\documentclass[twocolumn,showpacs,preprintnumbers,amsmath,amssymb,pra]{revtex4}

\usepackage{graphicx}
\usepackage{dcolumn}
\usepackage{bm}

\newcommand{\ket}[1]{\vert #1\rangle}
\newcommand{\bra}[1]{\langle #1\vert}
\newcommand{\e}{\mathrm{e}}
\newcommand{\ii}{\mathrm{i}}

\DeclareMathOperator{\Rb}{\mathrm{Rb}}

\begin{document}

\author{T. Calarco$^{1,2}$, U. Dorner$^{1}$, P. Julienne$^3$,
C. Williams$^3$, and P. Zoller$^{1}$ }
\affiliation{$^1$Institute for Theoretical Physics, University of Innsbruck, \\
and Institute for Quantum Optics and Quantum Information of the
Austrian Academy of Sciences, A-6020 Innsbruck, Austria\\
$^2$European Centre for Theoretical Studies in Nuclear Physics
and Related Areas, I-38050 Villazzano (TN), Italy\\
and CRS BEC-INFM, Dipartimento di Fisica, Universit\`a di Trento,
I-38050 Povo (TN), Italy\\
$^3$ National Institute of Standards and Technology, Gaithersburg,
MD 20899-8423, U.S.A. }

\title{Quantum computations with atoms in optical lattices:\\
marker qubits and molecular interactions}

\begin{abstract}
We develop a scheme for quantum computation with neutral atoms,
based on the concept of ``marker'' atoms, i.e., auxiliary atoms
that can be efficiently transported in state-independent periodic
external traps to operate quantum gates between physically distant
qubits. This allows for relaxing a number of experimental
constraints for quantum computation with neutral atoms in
microscopic potential, including single-atom laser addressability.
We discuss the advantages of this approach in a concrete physical
scenario involving molecular interactions.
\end{abstract}
\date{\today}

\maketitle
\section{\label{sec:sec1}Introduction}

Manipulation of cold atoms in microscopic traps is one of the
major highlights of the extraordinary progress experienced by
atomic, molecular and optical (AMO) physics over the past few
years, and has lead to important successes in the implementation
of quantum information processing \cite{cirac04}. By employing a
quantum phase transition it is possible to load large numbers of
neutral atoms in highly regular patterns within an optical lattice
\cite{SFMott}. This system is very promising both in terms of
quantum simulation of condensed matter physics, and more in
general of quantum information processing. Hence, over the last
few years, several implementations of neutral-atom quantum
computing, exploiting various trapping methods and entangling
interactions, have been proposed
\cite{charron02,tian03,pachos03,dorner03,rabl03,mompart03,
eckert02,lukin01,andersson01,ripoll03,brennen00,
brennen99,grangier01,dumke02,repeater,folman02,cote03}.

In this paper we study quantum computing with neutral atoms in
optical lattices based on the concept of  {\em marker} and {\em
messenger} atoms. We consider a situation where qubits are
represented by the internal longlived atomic states, and these
qubit atoms are stored in a (large) regular array of microtraps
realized by an optical lattice. These qubit atoms remain frozen at
their positions during the quantum computation. In addition to the
atoms representing the qubits, we consider an auxiliary ``marker
atom'' (or a set of marker atoms) which can be moved between the
different lattice sites containing the qubits. The marker atoms
can either be of a different atomic species or of the same type as
the qubit atoms, but possibly employing different internal states.
These movable atoms serve two purposes. First, they allow
addressing of atomic qubits by ``marking'' a single lattice site
due to the marker-atomic qubit interactions: this molecular
complex can be manipulated with a laser without the requirement of
focusing on a particular site. Second, the movable atoms play the
role of ``messenger'' qubits which allow to transport quantum
information between different sites in the optical lattice, and
thus to entangle distant atomic qubits.

The first key element in our scheme is the {\em transport of
marker (or messenger) atoms} in an off-resonant {\em
time-dependent superlattice}. By changing laser parameters with an
appropriate protocol we move the marker atoms from site to site
while leaving the qubit atoms frozen at their respective
positions. We note that to move a marker atom only the {\em
global} laser parameters generating the superlattice need to be
changed. In the case of several marker atoms on a lattice
(arranged e.g. in a certain spatial pattern) they will be moved in
parallel by these global lattice operations. The time scale for
these lattice movements can be of the order of the oscillation
period in the confining lattice potential. In addition, two
distinctive properties of the scheme are: (i) The superlattice can
be realized by a very far-offresonant optical lattice. Thus there
is no requirement for a qubit- (or spin-dependent) optical lattice
as in the case of collisional gates which in case of Alkali atoms
require tuning of the lattice laser between the excited atomic
fine structure states. This allows to strongly suppress
decoherence due to spontaneous emission in the present scheme.
(ii) There is significant freedom in choosing the internal atomic
states representing the qubits: in particular, we can choose
atomic states corresponding a ``clock transition''. These clock
states are insensitive to the (stray-) magnetic fields, again
improving decoherence properties of the atomic qubits. Also this
is in contrast to moving atoms in spin-dependent lattices for
collisional gates, where the qubit states are typically very
sensitive to magnetic fields.

A second key element is that we employ {\em resonant molecular
interactions} between marker and qubit-atoms, as provided by
magnetic or optical Feshbach resonances. This implies two features
of the present scheme: (i) Due to the resonant character combined
with the spatial confinement of atoms in the optical lattice,
these interactions can be comparable to the trap spacing in the
optical lattice, and thus the time scale of operations becomes of
the same order of magnitude as the one for the transport in the
lattice. (ii) In addition, these resonant molecular interactions
can be made internal state (qubit) dependent which gives a
mechanism for entangling the marker and atomic qubits, and to
perform swap operations of the atomic qubit to the marker atom.

The article is organized as follows: in Sec.~\ref{sec:sec2} we
introduce the general concept of quantum computing via ``marker''
qubits, and we specialize it to the case of atomic qubits in
optical lattices. In Sec.~\ref{sec:sec3} we develop and simulate a
procedure to effect selective atom transport in spin-independent
lattices. Sec.~\ref{sec:sec4} describes the theory of resonant
collisions in confined geometries, suitable for the treatment of
Feshbach resonance in tightly confining traps. Sec.~\ref{sec:sec5}
discusses the dynamics of one- and two-qubit operations using the
aforementioned ingredients. Conclusions are drawn in
Sec.~\ref{concl}.

\section{\label{sec:sec2}Concepts of quantum computing with ``marker'' atoms}

\subsection{General concept}
\begin{figure}[h!]
  \begin{center}
    \includegraphics[width=0.8\columnwidth]{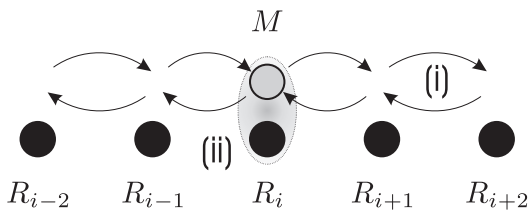}
    \caption{Basic operations with a marker atom on a quantum
    register: (i) forward and backward transport steps. (ii) local interaction with register qubits.}
    \label{concept}
  \end{center}
\end{figure}

The scheme we are introducing is based on a quantum register
formed by separately stored qubits, that never interact directly
with each other. To mediate entangling operations between
different register ($R$-type) qubits, we introduce ``marker'' or ``messenger''
($M$-type) qubits, that can be transported through between
different register locations. Direct coupling can only take place
between a register and a marker qubit. In the simplest situation
there is only one $M$ qubit present in a certain register location
$R_i$. Different operations are then possible (see
Fig.~\ref{concept}):
\begin{itemize}
\item[(i)] The $M$ qubit can be transported forward and backward throughout the string of register qubits thus being able to reach an arbitrary location $R_i$.

\item[(ii)] A local interaction between the $M$ and the $R$ qubit may
be activated to perform single and two qubit gates.
\end{itemize}
The role of the $M$ qubits in our
scheme is twofold. On one hand they will allow us to address
single register atoms without the need for addressing single
lattice sites, i.e. they act as a ``marker'' for a certain register atom.
On the other hand they act as information carrier
performing effective entangling operations between physically
distant register qubits. In this case they act as a ``messenger'' transporting quantum information.
However, to simplify language, in the following we will denote the $M$ qubits always as marker qubits or marker atoms.

\begin{figure}[h!]
  \begin{center}
    \includegraphics[width=\columnwidth]{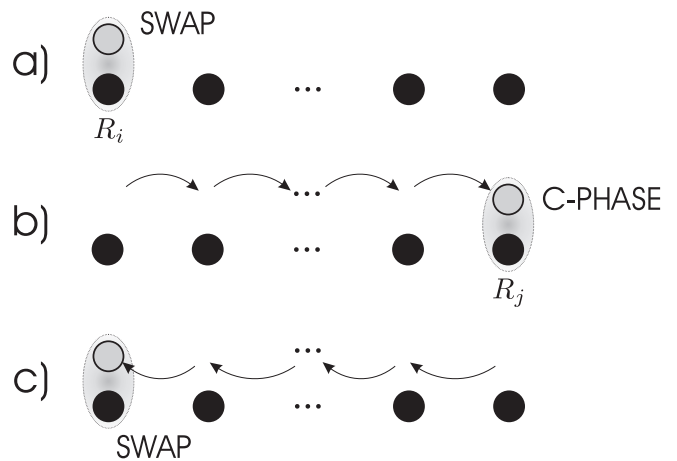}
    \caption{Realizing an entangling operation between
    distant atoms based on the elementary steps described in
    Fig.~\protect\ref{concept}: a) swapping the first qubit onto
    the marker atom; b) transporting the marker atom unto the
    second qubit and local interaction; c) transport back to the
    first qubit and inverse swap.}
    \label{distgate}
  \end{center}
\end{figure}

The interaction we apply depends on the logical state of both of the involved
qubits, i.e. it enables us to perform two-qubit gates like a
controlled-phase  or a swap gate between the marker and the
register atom. This in turn, combined with forward and backward transport operations (i), allows to construct sequences of operations that enact two-qubit gates between arbitrary register qubits. This works as
follows (see Fig.~\ref{distgate}):
\begin{itemize}
\item[a)] the state of qubit $R_i$ is first swapped onto the
marker $M$;

\item[b)] the marker is transported to location $j$, where a
two-qubit gate is performed between $M$ and $R_j$;

\item[c)] the marker is transported back to location $i$ and its
state is swapped back onto $R_i$.
\end{itemize}
At the end of the process, the marker qubit recovers its initial
state, while the net effect is that a gate operation has been
performed between register qubits $R_i$ and $R_j$. Beside logical
operations, this allows for the creation of distant entangled
(e.g., EPR) pairs that can be subsequently used for teleportation
between different quantum memory locations, for state purification
in error correcting protocols and for scalable probabilistic gates
\cite{duan04}.
\subsection{Implementation of the concept with atoms in periodic
trapping potentials}
In the following we want to briefly outline how the above general
concept can be implemented. A detailed description can be found in
Sec.~\ref{sec:sec3}-\ref{sec:sec5}.  As described above the two key
ingredients of our scheme are (i) the transport of marker atoms and
(ii) the application of a strong local interaction. We concentrate in
this work on an implementation with neutral atoms stored in a
two-component optical superlattice.  However, our scheme may be
transferred to other systems, including atom chips \cite{atomchip}.
We consider single atoms stored in the ground state of separate wells
which we model as a 1D periodic potential (see Sec.~\ref{sec:sec3a}), with a simple filling
pattern of one register atom $R_i$ every second lattice site. The
ground states of the remaining sites may (or may not) be occupied by
marker atoms, which can be of the same species as the register atoms,
and the tunnel coupling between neighboring sites is assumed to be
negligible, so that marker atoms do not interact with register atoms
unless the potential is modified. The quantum information is stored in
two appropriate internal states $\ket{0},\,\ket{1}$ of the atoms
(which will be specified in Sec.~\ref{sec:sec4}).

As described in Section \ref{sec:sec3}, the transport of the marker
atoms (i) is realized by globally changing the external lattice
control parameters which allows for creating a periodic array of
double-well structures with different well depths. In this way the
marker atom can be transferred from its initial site into the first
excited state of one of the neighboring wells, while the register atom
located there (as well as any other register atom in the lattice)
remains in its trap ground state. From here the marker can be
transported further to the ground state of the next site (which is not
occupied by an atom) or back again to its initial position. By
repeating these transport steps the marker atoms can be transported to
an arbitary lattice site. This scheme avoids there being two
permanently interacting atoms at any lattice site.

Due to the fact that the lattice parameters are changed globally, all
marker atoms undergo the same, parallel movement. Thus, when more
ground-state marker atoms are introduced at different sites, a certain
lattice transformation will transport all of them in the same way. By
suitably choosing the pattern of marker atoms, multi-qubit operations
can be carried out in parallel or with pre-defined patterns (an
example is encoding and syndrome extraction for error correction).

When a marker and a register atom are at the same site we realize the
coupling (ii) of Fig.~\ref{concept} by making use of the strong molecular interaction
between the marker and the register atom, which can be controlled
by an external magnetic field giving rise to a Feshbach resonance
\cite{Feshbach}. The physics behind this mechanism, as well as the gate
operations, will be detailed in Secs.~\ref{sec:sec4} and
\ref{sec:sec5}, respectively. Of course, this sort of interaction can be employed in
any neutral-atom quantum computation proposal. In this paper, we will
outline its general features, and we will focus on its specific use in the context of marker-atom quantum computing.

The principle of the single qubit gate is the following.  The marker
atom is transported to the register atom we want to address. Then the
molecular interaction is ``switched on'' via an external magnetic or
optical field, i.e. we perform a Feshbach ramp which leads to a level
splitting of the atomic states.  Clearly this splitting is only
present at the site with two atoms.  With appropriately detuned
external lasers we can then perform arbitrary single qubit rotations
\begin{align}
  \ket{0}\longrightarrow &\cos(\alpha)\ket{1} -\ii\sin(\alpha)\e^{\ii\varphi} \ket{0} \nonumber\\
  \ket{1}\longrightarrow &-\ii\sin(\alpha) \ket{1} +
  \cos(\alpha)\e^{\ii\varphi}\ket{0}
\end{align}
where the angle $\alpha$ is given by the interaction time with the
lasers (and their intensities) and the phase $\phi$ is determined by
the dynamics of the Feshbach ramp.  The lasers have not to be focused
down to the lattice constant. The spatial width is merely limited by
the distance to the next marker atom (except if we want to perform the
same rotation there).

The principle of two qubit gates is already shown in
Fig.~\ref{distgate} where in step $b)$ we either perform a swap operation between register and marker atom,
$\ket{\epsilon_1\epsilon_2}\rightarrow\ket{\epsilon_2\epsilon_1}$,
or a phase gate, $\ket{\epsilon_1\epsilon_2}\rightarrow
\exp[\ii\varphi(1-\epsilon_1)(1-\epsilon_2)] \ket{\epsilon_1\epsilon_2}$
where $\epsilon_{1,2}\in\{0,1\}$.
As we will describe in Sec.~\ref{sec:sec5}, the phase gate as well as the swap gate are again based on the tunable
molecular interaction between two atoms at one lattice site. In the
first case the phase is acquired by a Feshbach ramp which affects only
the state $\ket{00}$ while in the case of the swap gate we need,
similar as in the case of the single qubit rotation, an additional
laser field which couples resonantly the state which is shifted by the
molecular interaction and the states $\ket{01}$ and $\ket{10}$. The
laser will again affect only the sites where two atoms are present
thus again it does not have to be focused.


Beside relaxing addressability constraints, our scheme bears several
other advantages: for instance, it does not require a state-dependent
lattice \cite{jaksch99}.  As described in Sec.~\ref{sec:sec1} that
method has a couple of disadvantages and, furthermore, the realization
and stabilization of such potentials poses a major experimental
challenge. In our scheme the quantum register logical state never gets
entangled with the atomic motion, eliminating a major source of
decoherence. Even collisional phases, acquired by the marker atom
while being transported over occupied lattice sites, can be made
state-insensitive by an appropriate choice of the atomic hyperfine
states (a typical example being Rb, for which the singlet and the
triplet scattering lengths coincide), thus contributing only a global
phase to the evolution of the whole register.

\section{\label{sec:sec3}Atom transport in time dependent superlattices}
The transport scheme we describe in this section makes use of a
time dependent optical superlattice configuration which can be far-offresonant from the relevant optical transition to avoid
spontaneous emission and is not specialized to specific atomic
species.  The atom transport is independent of the considered
internal states and allows for using the $m=0$ states of different
hyperfine manifolds, i.e. states of a ``clock transition'' which
are not affected by external magnetic fields.

In the following we will describe the laser configuration which is
necessary to realize the superlattice and detail the transport of
single atoms in the periodic potential by changing the intensities
and phases of the lasers. We will furthermore discuss optimization
methods.
\subsection{\label{sec:sec3a}Realization of the superlattice}
\begin{figure}[t]
  \begin{center}
    \includegraphics[]{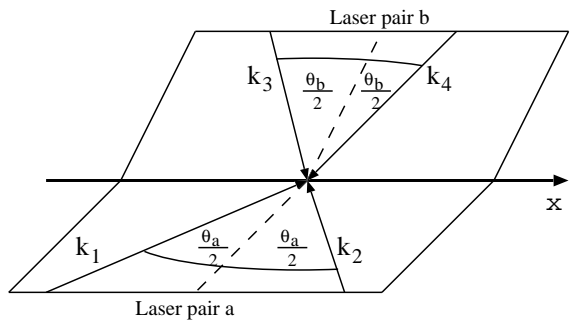}
    \caption{
      Laser setup. The plane containing the wavevectors $\vec k_{1,2}$
      of a laser pair $a$ intersects the plane containing the
      wavevectors $\vec k_{3,4}$ of a laser pair $b$ at the $x$-axis.
      The planes can have a finite angle between them. The angle
      between the lasers in each of the pairs are $\theta_a$ and
      $\theta_b$, respectively (see also \protect\cite{Rolston03}).}
    \label{fig_lasers}
  \end{center}
\end{figure}

For the realization of the superlattice potential we propose using a
configuration of four intersecting lasers like it was used in
\cite{Rolston03}. The setup is shown in Fig.~\ref{fig_lasers}.
Two pairs of laser beams intersect with an angle $\theta_a$ and $\theta_b$, respectively
 The lasers of frequency $\omega_a$ (pair a)
and $\omega_b$ (pair b) interact with atoms which are considered as
two level systems with transition frequency $\omega_0$.
In an interaction picture, the Hamiltonian for the internal degrees of
freedom of an atom can then be written as
\begin{align}
H = H_0 +
\hbar\Delta_a\sigma_+\sigma_-
- \frac{1}{2}\Big [&\sigma_+\vec d \left(\vec E_a(\vec r) + \vec E_b(\vec r)\e^{\ii\delta t}\right)
\nonumber\\
+&\sigma_-{\vec{d}}^*\left ({\vec E}_a^\dagger(\vec r) + {\vec E_b}^\dagger(\vec r)\e^{-\ii\delta t}\right) \Big ]
\end{align}
where $H_0={\vec p}^{\;2}/2m$ is the kinetic energy operator. Furthermore we introduced $\Delta_{a,b} = \omega_0-\omega_{a,b}$, $\delta =
\omega_a-\omega_b$, the atomic dipole moment $\vec d$ and the electric
fields
\begin{eqnarray}
\vec E_a(\vec r) &=& \sum_{j = 1,2} \mathcal{E}_j\vec\varepsilon_j\e^{\ii(\vec k_j\vec r + \phi_j)}, \\
\vec E_b(\vec r) &=& \sum_{j = 3,4}
\mathcal{E}_j\vec\varepsilon_j\e^{\ii(\vec k_j\vec r + \phi_j)},
\end{eqnarray}
where $\vec\varepsilon_j$ are the normalized polarization vectors,
$\mathcal{E}_j$ the field amplitudes and the wavevectors have the
magnitude $k_a\equiv \vert \vec k_1\vert = \vert \vec k_2\vert =
\omega_a/c$, $k_b\equiv \vert \vec k_3\vert = \vert \vec k_4\vert =
\omega_b/c$. \\
The laser pairs are assumed to be far off detuned from atomic
resonance and from each other so we can adiabatically eliminate
the upper atomic level and obtain an effective Hamiltonian in
position representation
\begin{equation}
\label{Heff}
H_{\text{eff}} = -\frac{\hbar^2}{2m}{\vec\nabla}^2 -\left( \frac{\vert \vec d\vec E_a(\vec r)
    \vert^2}{4\hbar\Delta_a} + \frac{\vert \vec d\vec E_b(\vec r)
    \vert^2}{4\hbar\Delta_b}\right ).
\end{equation}
According to the geometry of the laser setup the second term of Eq.~(\ref{Heff}) can be written as
\begin{align}
  U(x) &= -\Bigg[ \frac{1}{4\hbar\Delta_a}\sum_{j=1,2}\vert \vec
  d\vec\varepsilon_j\vert^2\mathcal{E}_j^2
  +\frac{1}{4\hbar\Delta_b}\sum_{j=3,4}\vert \vec
  d\vec\varepsilon_j\vert^2\mathcal{E}_j^2
  \nonumber\\
  & +\frac{(\vec d\vec\varepsilon_1)({\vec
      d}^*\vec\varepsilon_2) \mathcal{E}_1\mathcal{E}_2}
  {4\hbar\Delta_a} \cos\big(2k_ax\sin(\theta_a/2) + \phi_1-\phi_2\big) \nonumber\\
  &+\frac{(\vec d\vec\varepsilon_3)({\vec d}^*\vec\varepsilon_4)
    \mathcal{E}_3\mathcal{E}_4} {4\hbar\Delta_b}
  \cos\big(2k_bx\sin(\theta_b/2) + \phi_3-\phi_4\big)
  \Bigg]
  \label{laser_potential}\\
  &\equiv U_0 + U_1 \cos\big(2k_ax\sin(\theta_a/2) +
  \phi_1-\phi_2\big) \nonumber\\
  &\quad + U_2 \cos\big(2k_bx\sin(\theta_b/2) + \phi_3-\phi_4\big),\label{optpot2}
\end{align}
where $U_0$ is merely a constant. In Sec.~\ref{sec:sec3b} we will choose it according to Eq.~(\ref{U0}). If we set $\phi_1=\phi_2$,
$\phi\equiv \phi_4-\phi_3$, $\theta_a = \pi$ (counterpropagating
lasers) and
\begin{equation}
\theta_b = 2\arcsin\left(\frac{k_a}{2k_b}\right)
\approx 2\arcsin\left(\frac{1}{2}\right) = \frac{\pi}{3},
\end{equation}
the potential is of the type
\begin{equation}
\label{Ux}
U(x) = U_0 + U_1 \cos(2\kappa x) + U_2\cos(\kappa x-\phi).
\end{equation}
The potential (\ref{Ux}) leads to a particle confinement along the
$x$ axis.  We will assume in the following the confinement in
the transverse directions $y$ and $z$ to be much stronger than
along the $x$-direction so that we have effectively a one dimensional
system.

\subsection{\label{sec:sec3b}Single atom transport}
The transport of an atom through the lattice is achieved by varying
the amplitudes $U_i(t)$ and the relative phase $\phi(t)$ of the two
lattice components, which is done by changing the intensities and
phases of the lasers-see Eq.~(\ref{laser_potential}).  The potential (\ref{Ux})
becomes then time dependent: $U(x)\rightarrow U(x,t)$.
\begin{figure}[t!]
  \begin{center}
    \includegraphics[]{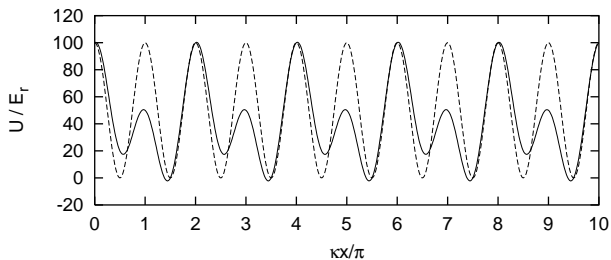}
    \caption{
      Potential $U(x,t)$ at two different times and $V=100\,E_r$ (with
      recoil energy $E_r = \hbar^2\kappa^2/2m$). At $t_0=0$
      $u_1(0)=u_2(0)=0$ (dashed line) while at $t_1>0$ we set
      $u_1(t_1)=0.5,\,u_2(t_2)=0.2$ and $\sigma=1,\,l=0$ (solid line).
      In the latter case the difference between the minima is
      approximately $20\,E_r$ and the height of every second barrier
      is reduced by approximately $50\%$.}
    \label{fig_potential}
  \end{center}
\end{figure}

For the description of the transport process it is useful to rewrite
the potential (\ref{Ux}) depending on two parameters $u_1(t)$ and
$u_2(t)$:
\begin{eqnarray}
  U_0(t) &=& \frac{V}{4}[2 - u_1(t) + u_2(t)],\label{U0}\\
  U_1(t) &=& \frac{V}{4}[2 - u_1(t) - u_2(t)],\\
  U_2(t) &=& \frac{V}{2}\sqrt{u_1^2(t) + u_2(t)^2},\\
 \phi(t) &=& \sigma\arctan{[u_2(t)/u_1(t) + l\pi ]}.
\end{eqnarray}
At time $t=0$ we set $U_2(t) = 0$, i.e. $u_1(0)=u_2(0)=0$ and thus we
simply have a cosine-potential of depth $V$ and periodicity
$a_0=\pi/\kappa$. However, in general the periodicity of the lattice is $a_1=2\pi/\kappa$ and the shape of the optical potential can be designed in the following way:

The parameter $u_1(t)$ controls approximately the height $V[1-u_1(t)]$
of every {\em second} barrier depending on the parameter
$l\in\{0,1\}$: If $l=0$ the height of every ``odd'' barrier is changed
while in case of $l=1$ the height of every ``even'' barrier is
changed, i.e. in the first case barriers with maxima at
$x=(2j+1)\pi/\kappa$ and in the latter case barriers with maxima at
$x=2j\pi/\kappa$ are modified for $U_2 = 0$ and $j\in\mathbb{Z}$. By changing
$u_1(t)$ we can thus create a specific periodic array of double well
potentials. The parameter $u_2(t)$ controls additionally the
difference of the minima~$Vu_2(t)$ of such a double well potential,
while $\sigma\in\{\pm1\}$ determines if the left ($\sigma=1$) or right
($\sigma=-1$) well is raised.

An example is shown in Fig.~\ref{fig_potential} for $l=0,\,\sigma=1$ and
two different values for $u_1$ and $u_2$. The potential $U(x,t)$ is
given in units of the recoil energy $E_r = \hbar^2\kappa^2/2m$ where
$m$ is the atomic mass appearing in the time dependent Schr\"odinger
equation
\begin{equation}
\label{schrod}
\ii\hbar\frac{d}{dt}\psi(x,t) = -\frac{\hbar^2}{2m}\frac{d^2}{dx^2}\psi(x,t) + U(x,t)\psi(x,t)
\end{equation}
which has to be solved for the study of single atom transport.
\begin{figure}[t]
  \begin{center}
    \includegraphics[]{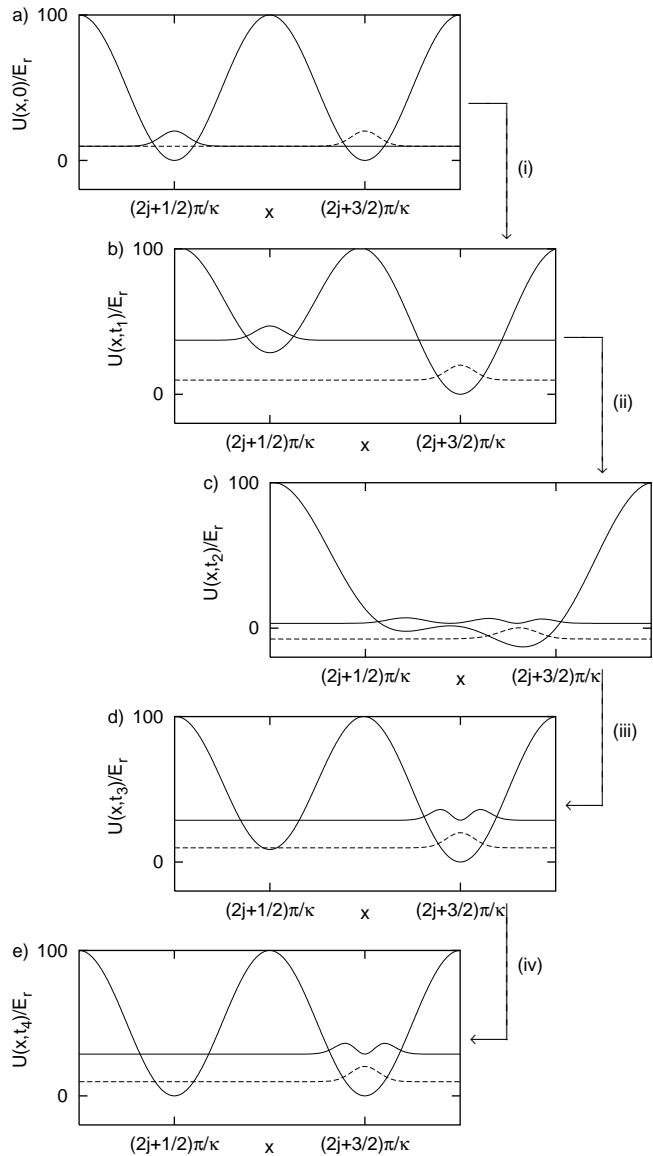}
    \caption{
      Illustration of steps (i)-(iv) of the transport process as
      described in the text. Shown are also the square of the absolute
      value of the single particle wavefunctions of the atom which is
      transported (i.e. the ``marker atom''), $\vert \psi^M(x)\vert^2$
      (solid lines), and of the register atom which is supposed to
      remain at its lattice site, $\vert \psi^R(x)\vert^2$ (dashed
      lines). The example shown requires $\sigma=1$ and $l=0$.}
    \label{fig_sequence}
  \end{center}
\end{figure}
The elementary steps of the atom transport are done by tunnelling
in the double well potentials. An example is shown in
Fig.~\ref{fig_sequence} for $\sigma=1$, $l=0$ and $V=100\,E_r$:

At the initial time $t=0$, where $u_1(0)=u_2(0)=0$, we consider two
neighboring wells, with one atom in the motional ground state of the
left well (Fig.~\ref{fig_sequence}a). The probability densities
$\vert\psi^M(x,t)\vert^2$ of the atom are indicated by the solid lines
in this figure. The superscript $M$ indicates the wavefunction of the
atom to be transported, i.e. the wavefunction of the ``marker atom''
as introduced in Sec~\ref{sec:sec2}. Also shown in this figure by the
dashed lines are the probability densities of the ``register atom''
$\vert\psi^R(x,t)\vert^2$, initially located in the ground state of
the right well in this example and which is supposed to remain at its
lattice site during the transport.
Our goal in the process described here is to transfer the left atom into the
first excited state of the right well without affecting the other one.
This can be accomplished by changing the parameters $u_{1,2}(t)$
according to the following steps, which are illustrated in
Fig.~\ref{fig_sequence}:
\begin{itemize}
\item[(i)] Between the times $t=0$ and $t=t_1$ we raise very rapidly
  the minimum of the left well, such that its ground state crosses in energy the
  right well's first excited state.
\item[(ii)] In the time interval $[t_1,t_2]$ we lower the central
  barrier down to a point where the atom can tunnel from left to right
  while at the same time we start to lower the left well.
\item[(iii)] In the time interval $[t_2,t_3]$ the barrier is raised up
  again while we continue to lower back the left well.
\item[(iv)] During the time interval $[t_3,t_4]$ we restore the
  initial potential shape.
\end{itemize}
By doing steps (ii) and (iii) adiabatically the atom stays in the
second excited state of the double well potential which is at $t_1$
the ground state of the left well and at $t_3$ the first excited state
of the right well. Thus the atom is transported from left to right.

An effective transport procedure requires appropriate ``pulse
functions'' $u_1(t)$ and $u_2(t)$ while the direction of the transport
is governed by the parameters $\sigma,l$. Let us assume for simplicity
that the marker atom is located initially at a site $\kappa x_j =
(2j+1/2)\pi,\,j\in\mathbb{Z}$. Then the process shown in
Fig.~\ref{fig_sequence} requires $\sigma=1$ and $l=0$ while for a further
movement to the motional ground state on the right we have to set
$\sigma=-1$ and $l=1$ and to perform the same pulse functions {\em
  backwards in time}. For moving an atom from the ground state to the
first excited state of the {\em left} neighboring well we have to
set $\sigma=-1$ and $l=1$ and to apply the forward pulse functions.
In this case a further movement to the left ground state requires
$\sigma=1$ and $l=0$ and the backward pulse function. Note that
the abrupt changes of the phase $\phi$ take place while $U_2$ is
zero, i.e. when the corresponding lasers are completely blocked
off from the atoms. Every transport of an atom across the lattice
can be divided into these four elementary processes. Since the
pulse sequence is in all cases the same (except for time reversal)
we can focus in the following on the example shown in
Fig.~\ref{fig_sequence}.

The feasibility of our quantum computing scheme depends on the time
scale on which quantum operations can be performed. Clearly the latter
is directly connected to the speed of the transport process.  In this
respect steps (i) and (iv) can be performed over much shorter times
than step (ii) and (iii), which are limited for example by the energy
difference to the other motional states. In order to examine adiabatic
transport during step (ii) and (iii) it is thus necessary to study the
instantaneous eigenenergies of an atom during the transport process in
dependence on the parameters we can control, i.e. the pulse functions
$u_{1,2}(t)$.

Since $U(x,t)$ is periodic, we can calculate the instantaneous
eigenenergies and eigenfunctions of the single particle Hamiltonian by
introducing Bloch functions $\psi_k^{(n)}(x)=e^{\ii kx}u_k^{(n)}(x)$
with Bloch vector $k$ and band index $n$.  In Fourier space the
stationary Schr\"odinger equation takes the form
\begin{align}
\label{fourier_schrod}
\frac{\hbar^2}{2m}(q-k)^2\tilde u_k^{(n)}(q,t)&+\sum_{q'}\tilde U(q-q',t)\tilde u_k^{(n)}(q,t) \nonumber\\
&= E_k^{(n)}(t)\tilde u_k^{(n)}(q,t),
\end{align}
where
\begin{align}
  &\tilde u_k^{(n)}(q,t) = \frac{1}{a_i}\int_0^{a_i} dx\,\e^{\ii qx}u_k^{(n)}(x,t),\nonumber\\
  &u_k(x,t) = \sum_q \e^{-\ii qx} \tilde u_k(q,t),\quad q = \frac{2n\pi}{a_i},\,n\in\mathbb{Z}
\end{align}
and
\begin{align}
  \tilde U(q,t) = U_0(t)\delta_{q,0}
  + &\frac{U_1(t)}{2}(\delta_{2\kappa,q}+\delta_{2\kappa,-q}) \nonumber\\
  + &\frac{U_2(t)}{2}(\e^{\ii\phi(t)}\delta_{\kappa,q}+\e^{-\ii\phi(t)}\delta_{\kappa,-q}).
\end{align}
where $a_i$ with $i=0,1$ are the lattice constants, i.e.
$a_0=\pi/\kappa$ if $U_2=0$ and $a_1=2\pi/\kappa$ if $U_2\ne0$. By
assuming periodic boundary conditions the Bloch vector $k$ gets
quantized, i.e. $k=2n\pi/Ma_0,\,n=-M/2,\ldots ,M/2$ where $M$ is the
number of lattice sites.  Given the functions $u_k(x)$ we can
furthermore construct Wannier functions which are localized at
lattice sites $x_i$,
\begin{equation}
\label{wannier}
w^{(n)}(x-x_i) = \frac{1}{\sqrt M}\sum_k \e^{\ii k(x-x_i)}u_k^{(n)}(x).
\end{equation}
These functions are needed for example as initial states for solving
the time dependent Schr\"odinger equation~(\ref{schrod}). Since in our
considerations the lowest bands are always practically flat, i.e. the Bloch states
for a given band are approximately degenerate, the Wannier functions
are also in good approximation eigenstates of the Hamiltonian and thus
dispersion of the wave packet during the time evolution is negligible.
\begin{figure}[t]
  \begin{center}
    \includegraphics[]{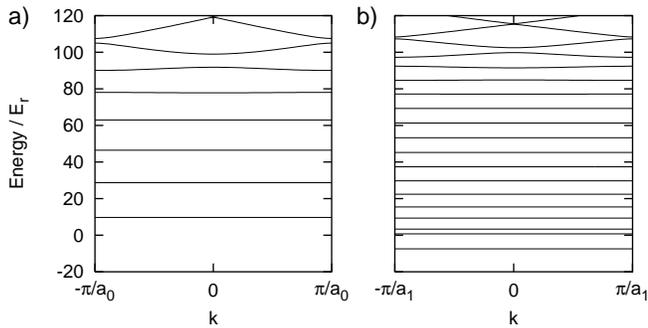}
    \caption{Band structures for
      two different values of $u_1,u_2$ and
      $V=100E_r,\,\sigma=1,\,l=0$. (a) Initial values of the lattice
      parameters ($u_1(0)=u_2(0)=0$). (b) Band structure during the
      transport process ($u_1=0.99,\,u_2=0.13$). The two figures
      correspond to the situations shown in Figs.~\ref{fig_sequence}a
      and \ref{fig_sequence}c.}
    \label{fig_bands}
  \end{center}
\end{figure}

Equation~(\ref{fourier_schrod}) is a linear system of equations which
can be solved numerically after truncating $q$ at sufficiently high
values. The instantaneous eigenenergies $E_k^{(n)}(t)$ gained in this
way are very important to find an adiabatic passage for the atom
transport.  An example for the band structure is shown in
Fig.~\ref{fig_bands}. As can be seen from this figure, $V$ is
sufficiently large to be in a tight binding regime, i.e. the lower
bands are flat and there is no tunnelling between different wells.

\begin{figure}[t]
  \begin{center}
    \includegraphics[width=\columnwidth]{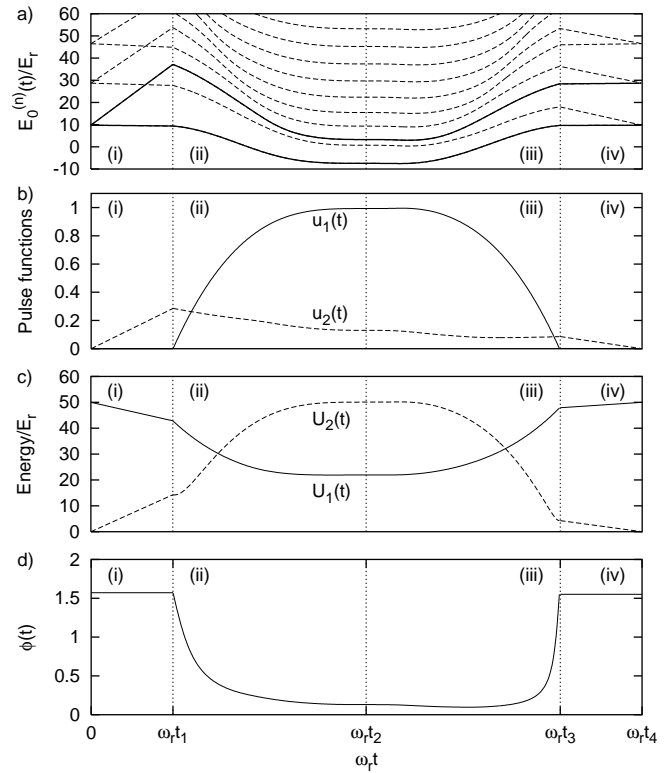}
    \caption{Adiabatic atom transfer between lattice sites:
      (a) Instantaneous eigenenergies $E^{(n)}_{k=0}(t)$. The upper
      solid line corresponds to the motional energy of the atom to be
      transported and the lower solid line corresponds to the register
      atom initially in the right well (cf. Fig.~\ref{fig_sequence}).
      (b) Corresponding control parameters $u_{1,2}$. (c) Amplitudes
      of the lattice components. (d) Phase difference between lattice
      components. The time axis is given in units of the recoil
      frequency $\omega_r\equiv E_r/\hbar$. The time interval $[t_1,t_3]$
       is not of the same scale as $[0,t_1],[t_3,t_4]$.}
    \label{transport}
  \end{center}
\end{figure}

For an efficient adiabatic transport the pulse functions have now to be
chosen such that the corresponding energies $E_k^{(n)}(t)$ behave in
an appropriate way during the transport steps (ii) and (iii), i.e. one
should for example avoid level crossings of the initial energy with
the energies of other states. An example is shown in
Fig.~\ref{transport}a.

As already mentioned the Bloch states in the lowest bands are
almost degenerate so we can restrict ourselves to a an arbitrary
value of $k$, e.g. $k=0$. The upper solid line in this figure
corresponds to the ``path'' of the atom to be transported
initially located in the left well while the lower solid line
indicates the path of the atom located in the right well. The
initial depth of the potential wells is $V=100\,E_r$ leading to a
``trap frequency'' in a single well of about $20\,\omega_r$
($\omega_r\equiv E_r/\hbar$ is the recoil frequency). Keeping the
height of the barrier constant we raise the minimum of the left
well such that the ground state energy of the left well crosses
the first excited state energy of the right well [step (i)]. Then
we proceed according to steps (ii) and (iii). By reducing the
height of the central barriers the trap frequency decreases and by
adjusting appropriately the pulse functions we avoid that the
solid lines cross the dashed lines. After time $t_3$ the original
potential shape is restored [step (iv)]. The pulse functions
$u_{1,2}(t)$ used for this example and the corresponding time
dependent physical relevant parameters $U_{1,2}(t),\,\phi(t)$ are
shown in Fig.~\ref{transport}b and Figs.~\ref{transport}cd,
respectively.


The fidelities of the processes, given by
\begin{equation}
F^A
=\left\vert \int dx\, {\psi^A}^*(x,T) \psi^A_\text{fin}(x) \right\vert^2,
\end{equation}
are numerically calculated by solving the time dependent single
particle Schr\"odinger equation in position representation
(\ref{schrod}) by using the Crank-Nicholson scheme \cite{NR} where
as initial state $\psi^A_\text{ini}$ and final state
$\psi^A_\text{fin}$ we choose Wannier functions
(\ref{wannier}) which are located in the corresponding wells.  The
superscript $A\in\{M,R\}$ indicates again the wavefunction
of the atom to be transported (marker atom) and the atom which is
supposed to stay located at its well (register atom).

In case of the example of Fig.~\ref{transport} we get $F^M =
99.91\%$ for propagating the marker atom wavefunction and $F^R =
99.98\%$ for propagating the register atom wavefunction from $t_1$ to
$t_3$ in a time $T=t_3-t_1 = 20/\omega_r$. In the case of Rubidium
($\omega_r = 2\pi\times 3.8\,\mathrm{kHz}$) this would correspond to a
time $T=0.8\,\mathrm{ms}$ and for sodium ($\omega_r = 2\pi\times
25\,\mathrm{kHz}$) we would have $T=130\,\mu\mathrm{s}$.  For the
optical superlattice described in \cite{Rolston03} a laser power of $P
= 3\,\mathrm{mW}$ was sufficient to create a maximal potential depth
of $2U_i<60E_R$ for $\sideset{^{87}}{}\Rb$. Keeping the ratio of laser
intensity and detuning constant we have $U_i\sim \sqrt{P}$. For a
potential depth of roughly $U_i<50E_R$ which is required in the above
example (see Fig.~\ref{transport}c) we can estimate the required
maximal laser power to be merely $P\approx 8-9\,\mathrm{mW}$.

\subsection{Pulse optimization}
If we relax the constraint of adiabatic transport during step (ii) and (iii) the process
described in the previous subsection can be significantly accelerated.
In this case the pulse sequences have to be engineered in a certain
way which can be done by using quantum optimal control techniques as
detailed, e.g., in \cite{rabitz,hohenester,Sklarz02}. Thereby the
evolution of a quantum system governed by a set of control parameters (in our case these are the functions $u_{1,2}(t)$)
is tailored to reach a pre-determined target state $\psi_\text{fin}$ with optimized
fidelity within a specific time $T$. For notational convenience we will denote in the following the time $t=t_1$ as $t=0$ and  $t=t_3$ as $t=T$.

The basic idea is to minimize the
infidelity of the process with the constraint that the Schr\"odinger
equation has to be fulfilled. This amounts to find the stationary
point of a functional, leading to a set of equations for the
wavefunction and auxiliary states $\chi^A$ which are introduced
as Lagrange multipliers. In our case this functional takes the form
\begin{align}
&\mathcal{L}(\psi^M,\psi^R,\dot\psi^M,\dot\psi^R,\chi^M,\chi^R,u_1,u_2) = \nonumber\\
&\qquad\sum_{A\in\{M,R\}}\Bigg[
  1 - \left\vert \int dx\, \psi^A_{\mathrm{fin}}(x){\psi^A}^*(x,T) \right\vert^2
\nonumber\\
  &\qquad +2\mathrm{Re}\Bigg( \int_0^T dt \int dx\, \chi^A(x,t)\big(\dot\psi^A(x,t) \nonumber\\
  &\qquad\qquad\qquad +\frac{\ii}{\hbar}
  H(u_1(t),u_2(t))\psi^A(x,t)\big)^* \Bigg) \Bigg],
\label{functional}
\end{align}
with
\begin{equation}
\label{hamil}
H(u_1(t),u_2(t)) =  -\frac{\hbar^2}{2m}\frac{d^2}{dx^2} + U(x,u_1(t),u_2(t)),
\end{equation}
where the last term in Eq.~(\ref{hamil}) is the potential (\ref{Ux}), which
depends on the control parameters.  As can be seen from
Eq.~(\ref{functional}) we are looking for a minimum of the sum of the
infidelities of the process for the marker atom and the register atom.
Setting the derivatives with respect to
the arguments of $\mathcal{L}$ equal to zero leads to the following
set of equations,
\begin{align}
\label{schrod1}
&\ii\hbar\dot \psi^A(x,t) = H(u_1(t),u_2(t))\psi^A(x,t), \quad A\in\{R,M\},\\
\label{eq_chi}
&\ii\hbar\dot\chi^A(x,t) = H(u_1(t),u_2(t))\chi^A(x,t),\quad A\in\{R,M\}
\end{align}
with conditions
\begin{align}
\label{initial}
&\psi^A(x,0) = \psi^A_{\text{ini}}(x), \\
\label{end_value}&\chi^A (x,T) \equiv \psi_\text{fin}^A(x) \int dx\,\psi_\text{fin}^A(x) {\psi^A}^*(x,T),
\end{align}
and
\begin{equation}
\label{gradient}
0 = -2\mathrm{Im} \sum_{A\in\{M,R\}} K_j^A(u_1(t),u_2(t)),\quad j=1,2
\end{equation}
with
\begin{align}
  K_j^A(u_1&(t),u_2(t)) \equiv \nonumber\\
  &\int dx\, \psi^A(x,t) \frac{\partial
    U(x,u_1(t),u_2(t))}{\hbar\partial u_j(t)} {\chi^A}^*(x,t).
\end{align}
These equations are the basis of the optimal control algorithm which
minimizes $\mathcal L$ with respect to $u_j$.
Thereby we solve Eq.~(\ref{schrod1}) and (\ref{eq_chi})
numerically by introducing a discretized time axis with time step
$\Delta t$ and by using the Crank-Nicholson scheme. For the sake
of completeness we briefly describe the algorithm we use here.
The following procedure, called immediate feedback control, is
guaranteed to give a fidelity improvement at each iteration \cite{sola}.

The Schr\"odinger equations (\ref{schrod1}) are integrated from $t=0$ to $t=T$
leading to $\psi^A(x,T)$ with an initial guess for the control parameters
$u_{1,2}^{(0)}(t)$. At this point an iterative algorithm starts during which the controls
$u_{1,2}^{(n)}(t)$ are updated.

Let us assume that we are in the $n$-th iteration.  Taking the controls
$u_{1,2}^{(n)}(t)$, Eqs.~(\ref{eq_chi}) have to be solved backwards in
time, i.e. from $t=T$ to $t=0$, with ``end values'' (\ref{end_value})
which can be interpreted as the part of $\psi^A(x,T)$ that has
reached the objective. Given the solutions $\chi^A(x,0)$ the
functions $\chi^A(x,t)$ and $\psi^A(x,t)$ (with initial
conditions (\ref{initial})) are now again evolved forward in time
while the control parameters are updated during each time step
according to
\begin{align}
\label{upgrade}
u_j^{(n+1)}(t) = u_j^{(n)}(t) 
+\frac{2}{\lambda(t)}\text{Im}\sum_{A\in\{M,R\}} K_j^A(u_1^{(n)}(t),u^{(n)}_2(t)). \nonumber\\
\end{align}
During the forward evolution $\chi(x,t+\Delta t)$ is calculated using
the controls $u_{1,2}^{(n)}(t)$ while $\psi(x,t+\Delta t)$ is evolved
according to $u_{1,2}^{(n+1)}(t)$.  The weight $\lambda(t)$ is used to
enforce fixed initial and final conditions on the control pulses.
Given these solutions we go on with the next iteration.

\begin{figure}[t]
  \begin{center}
    \includegraphics[]{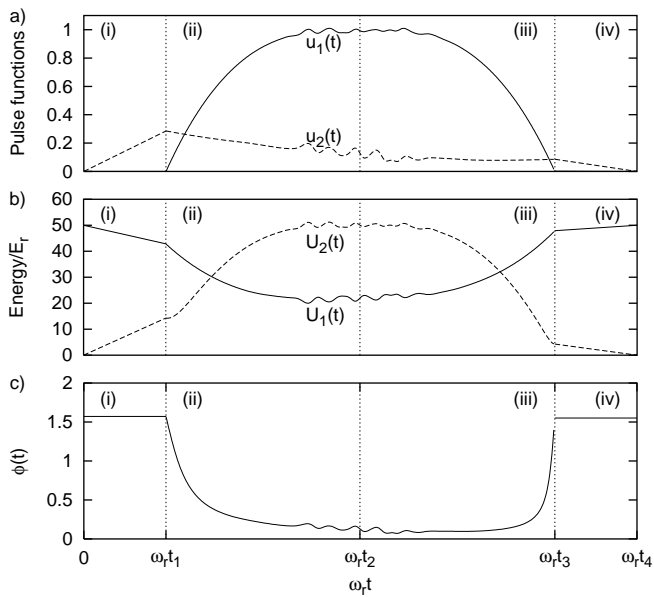}
    \caption{Non-adiabatic atom transfer between lattice sites:
      (a) Control parameters $u_{1,2}$. (b) Corresponding amplitudes
      of the lattice components. (c) Phase difference between lattice
      components. The time axis is given in units of the recoil
      frequency $\omega_r\equiv E_r/\hbar$. The time interval $[t_1,t_3]$
       is not of the same scale as $[0,t_1],[t_3,t_4]$.}
    \label{fig_wiggly}
  \end{center}
\end{figure}

The results of such a calculation after $135$ iterations are shown
in Fig.~\ref{fig_wiggly} which corresponds to the situation of
Fig.~\ref{fig_sequence}: Fig.~\ref{fig_wiggly}a shows the control
parameters for transferring the marker atom to the first excited
state of its right neighboring site while keeping the register
atom at its initial location.  Fig.~\ref{fig_wiggly}bc shows the
corresponding physical parameters. As starting values
$u^{(0)}_{1,2}(t)$ we took the pulses of the adiabatic example (see
Fig.~\ref{transport}).  The use of the optimized pulses leads to a
reduction of the transport time down to $T_3=5\hbar/E_R$, which corresponds to
$200\,\mu\mathrm{s}$ for rubidium and to $32\,\mu\mathrm{s}$ for
sodium with a fidelity of $F^M=F^R = 99.99\%$.

\section{\label{sec:sec4}Coherent resonant collisions in a trap}
The coupling scheme we are proposing can be
implemented either in dipole-force potentials, like optical
lattices \cite{optlatt}, or in static electromagnetic traps, like
atom chips \cite{atomchip}. Performing gate operations as
described in Sec.~\ref{sec:sec4} requires a strong molecular
interaction between register atom and marker atom. Atoms can be
coupled to molecular states either by means of Feshbach resonances
\cite{Feshbach} or through Raman photoassociation laser pulses
\cite{photoassociation}. For the sake of concreteness, we focus
here on Feshbach resonances in optical lattices -- however, all of
our arguments can be adapted, e.g., to Raman photoassociation on
atom chips. We consider $^{87}$Rb atoms trapped in a two-component
optical lattice (see, e.g., \cite{Rolston03}).

\subsection{Feshbach resonances in confined geometry}
\begin{figure}[t]
  \begin{center}
    \includegraphics[width=0.8\columnwidth]{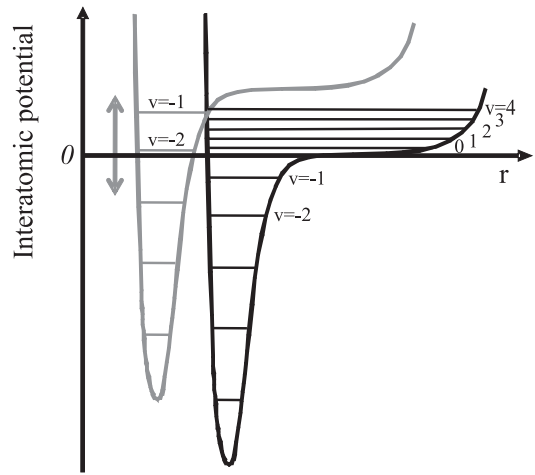}
    \caption{Interaction of
    two atoms in an external trap. The two curves schematically
    describe the Born-Oppenheimer potential for two
    scattering channels around a Feshbach resonance.}
    \label{bornoppen}
  \end{center}
\end{figure}

A schematic picture of the Born-Oppenheimer potential describing
their interaction in the relative coordinate $r$ is shown in
Fig.~\ref{bornoppen}.
Negative values of the excitation number $v$ label bound molecular
eigenstates of the dimer system, while positive $v$ values denote
unbound trapped two-atom states. Such a potential exists for each
collision channel $|\beta\rangle$, corresponding to the
relative-motion and hyperfine angular momentum quantum numbers of
the two colliding atoms. Feshbach resonances occur when a bound
state $|n_\beta\rangle$ ($v=-2$ in the example shown) crosses the
dissociation threshold for a state having the same quantum numbers
\cite{Feshbach} while changing an external magnetic field $B$.
Close to resonance, the scattering length varies as
\begin{equation}
A_\beta(B)=A_{bg}\left( 1-\frac{\Delta _\beta}{B-B_\beta}\right) ,
\end{equation}%
where $A_{bg}$ is a non resonant background scattering length, $B_\beta$ is the resonant magnetic field, and $\Delta _\beta$ is the width of the resonance.
The resonance energy varies almost linearly with the field
\begin{equation}
\varepsilon_\beta(B)=s_\beta(B-B_\beta), \label{resenergy}
\end{equation}%
with a slope $s_\beta$. We are interested in the dynamics of such a system in a confined
geometry. Following \cite{Mies00}, we shall model it by the
effective Hamiltonian
\begin{equation}
H_\beta=\varepsilon_\beta(B)|n_\beta\rangle\langle
n_\beta|+\sum_v(v\hbar\nu|v\rangle\langle
v|+V_v^\beta|v\rangle\langle n_\beta|+{\rm h.c.}),
\end{equation}
where the $|v\rangle$'s are the trapped relative-motion atomic
eigenstates of an isotropic harmonic oscillator trap having
frequency $\nu$. The couplings to the resonance are
\begin{equation}
\label{couplings} V_v^\beta=2\hbar\nu
\sqrt{\sqrt{4v+3}\;a_{bg}\delta_\beta/\pi}
\end{equation}
with $a_{bg}\equiv A_{bg}\sqrt{m\nu/\hbar}$,
$\delta_\beta\equiv\Delta_\beta s_\beta/(\hbar\nu)$. In a
different geometry, for instance in an elongated trap
characterized by a ratio $\gamma$ between the ground level
spacings in the transverse and in the longitudinal potential, the
couplings can be calculated by projection on the corresponding
eigenstates (see Appendix~\ref{app:cigar}). Accurate values for
the resonance parameters $\Delta_\beta$ and $B_\beta$, as well as for $A_{bg}$, are now available from both theoretical calculations and
recent measurements \cite{Verhaar02}.

\begin{figure}[t]
  \begin{center}
    \includegraphics[width=0.8\columnwidth]{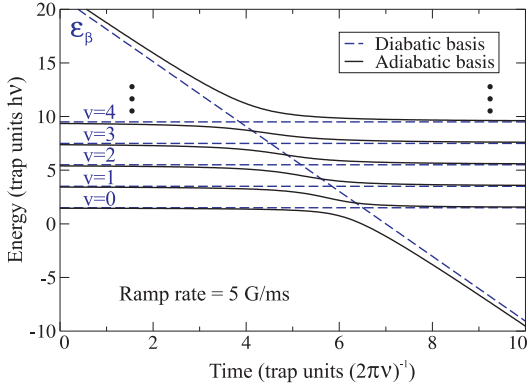}
    \caption{Adiabatic (solid) and diabatic (dashed) energies of a
    6-level expansion (molecular resonance level plus 5 trap levels
    of an isotropic 100 kHz trap) with a linear ramp of magnetic field with
    slope 5 G/ms. The adiabatic energy levels show a set of avoided crossings.
    We use the Feshbach resonance level near 100 mT. The coupling matrix element
    $V_0$is 0.884 $h\nu$ for this trap.}
    \label{diabatic}
  \end{center}
\end{figure}

The possibility of controlling the resonance energy via an
external magnetic field, as described by Eq.~(\ref{resenergy}),
provides a straightforward way to steer the interaction between
the atoms. Not only can the scattering length be varied over a
significant range -- the atoms can also be adiabatically coupled
into a molecular state. An example of this sort of process is
shown in Fig.~\ref{diabatic}. Here, the eigenvalues of the
interacting Hamiltonian are shown for a six-level model including
the five lowest unbound trap states plus the resonant state. The
latter is ramped across threshold by applying an external magnetic
field having a linear dependence on time. Both the so-called
``diabatic'' energies (i.e. those obtained by neglecting the
couplings to the resonance) and the adiabatic ones (i.e. the
actual eigenvalues of the full coupled Hamiltonian) are plotted
against time for a certain ramp rate. The important point to
notice is that the ground state of the relative motion is
adiabatically connected to the resonant state. Therefore, if the
atoms are prepared in their relative-motion ground state and the
resonance state is ramped across threshold from above, the atoms
are transferred into the bound state, whose energy depends on the
magnetic field - and the process is actually reversible. This
mechanism has been used for the creation of molecules in ultracold gases
\cite{heinzen02,ketterle03,wieman02,regal03,grimm03,salomon03,rempe04}, and
is very relevant in the present context of quantum information
processing due to its inherent state-dependent nature. Indeed, the
coupling to a specific resonant state $|n_\beta\rangle$ is only
effective for a particular entrance channel $|\beta\rangle$, while
in general all other combinations of atomic hyperfine states (that
is, of logical qubit states in our case) will be unaffected by the
resonance. Thus the resonance-induced energy shift will cause a
two-particle phase to appear only for that particular two-qubit
computational basis state. We will see in the next Section how to
use this effect in order to achieve a desired C-phase gate.

\subsection{Choosing qubit logical states}
We will identify our qubit logical states with the
clock-transition states
\begin{equation}
\left\vert 0 \right\rangle \equiv \left\vert
F=1,m_{F}=0\right\rangle,\quad
\left\vert 1 \right\rangle \equiv \left\vert
F=2,m_{F}=0\right\rangle
\end{equation}%
and the auxiliary state as
\begin{equation}
|x\rangle\equiv|F=1,m_F=1\rangle.
\end{equation}
The main advantage of this choice is that the qubit states are not
sensitive to the magnetic field, and hence not subject to
decoherence due to its fluctuations.

\begin{figure}[h!]
  \begin{center}
    \includegraphics[width=0.8\columnwidth]{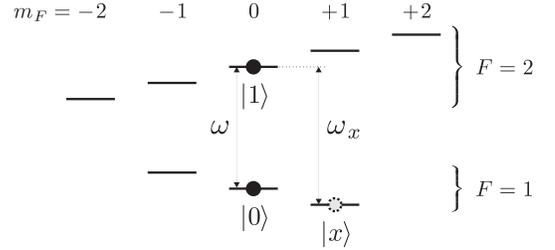}
    \caption{Internal level scheme for a single atom: qubit and auxiliary
    states.}
    \label{singlevels}
  \end{center}
\end{figure}

\begin{figure}[h!]
  \begin{center}
    \includegraphics[width=0.8\columnwidth]{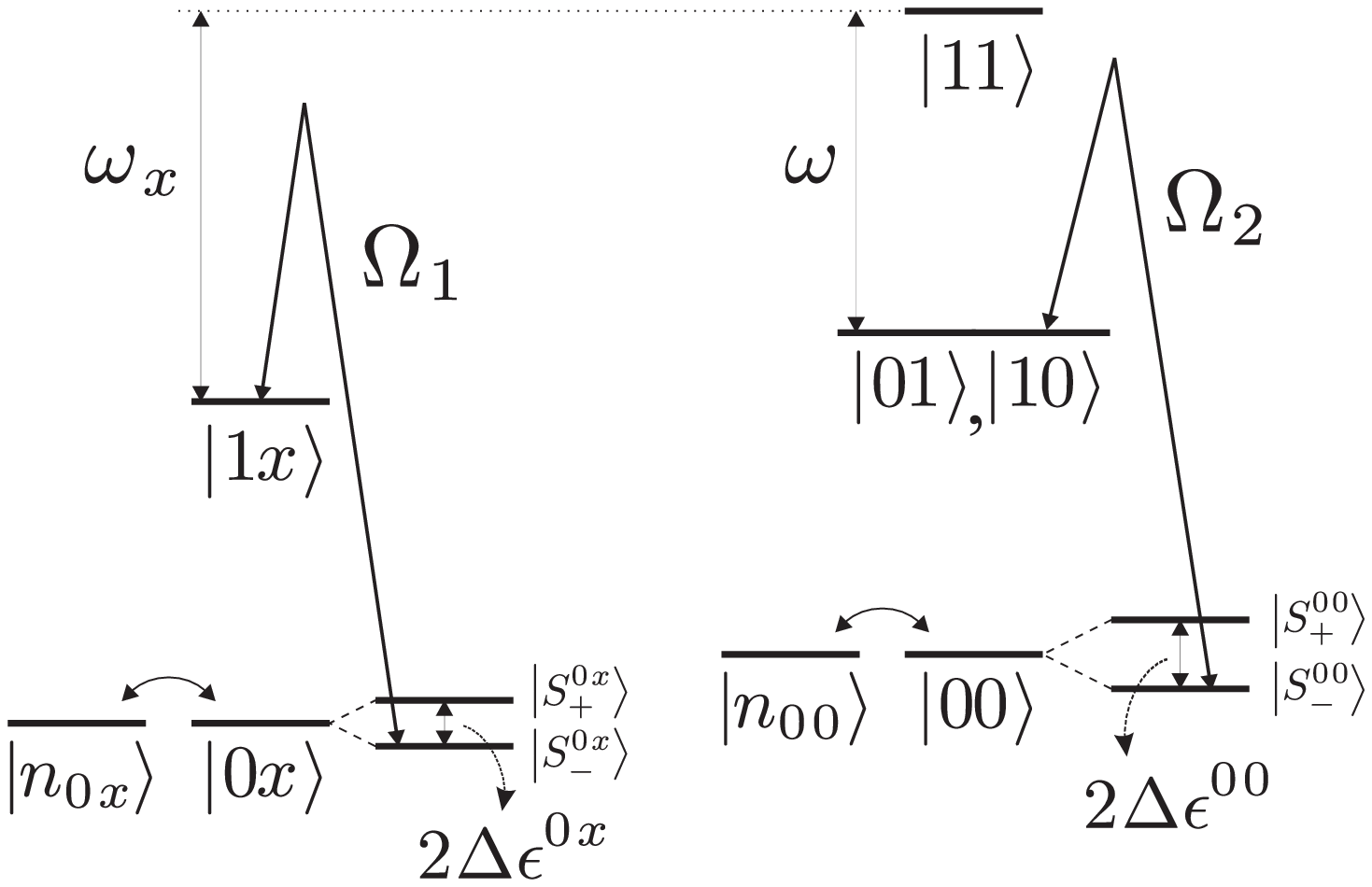}
    \caption{Internal level scheme for two coupled atoms:
    levels involved in single- (left) and in two-qubit  operations (right).
    The Raman transitions $\Omega_1$ ($\Omega_2$)
    used for single- (two-)qubit operations are shown.}
    \label{doublevels}
  \end{center}
\end{figure}

The level scheme for a single atom is shown in
Fig.~\ref{singlevels}.
When we consider two atoms, the relevant level scheme is described
by Fig.~\ref{doublevels}, if they occupy their relative-motion
ground state. Appendix~\ref{app:levels} shows that this is indeed
the case for two bosons stored in the ground and first axial
excited state of a cigar-shaped harmonic trap.
In Fig.~\ref{doublevels} the resonant levels $|n_\beta\rangle$ are
also shown, which are used to induce energy shifts for the purpose
of gate operation as described in the next Section. Indeed, in a
confined geometry, the coupling to such molecular states can
induce dressing of the trapped eigenstates with a half-splitting
$\Delta\epsilon^\beta$ --~controllable, by varying the external
field, up to a maximum value equal on resonance to the interaction
strength $V^\beta$~--, as shown in Fig.~\ref{doublevels} for the
collisional channels $|\beta\rangle=|00\rangle,|0x\rangle$.

\begin{figure}[t]
  \begin{center}
    \includegraphics[width=7truecm]{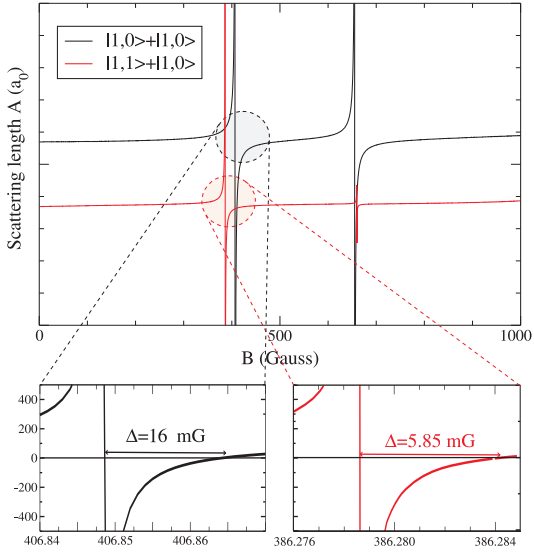}
    \caption{Dependence of the scattering length of $^{87}$Rb on the
    external magnetic field for collisions in channels $\ket{00}$ and $\ket{0x}$.}
    \label{resoscan}
  \end{center}
\end{figure}

Our calculation with a realistic molecular interaction potential
yields, among others, two resonances at $B_{0x}=386$ G and
$B_{00}=407$ G, having widths $\Delta_{0x}=5.7$ mG and
$\Delta_{00}=16$ mG, as shown in Fig.~\ref{resoscan}. These will
be employed in the following to effect logical gate operations.

\section{\label{sec:sec5}Quantum operations}
Let us now examine in detail how to use the features described
above in order to perform quantum computation in our system.
Preparing all atoms in an initial state by optical pumping
requires no single-qubit addressing and can be performed with
standard techniques. As the next step, according with the above
discussion, performing a specific algorithm will require a certain
pattern of marker atoms. These can be prepared either in a
periodic fashion, by means of a superlattice tuned to the
appropriate transition, or in an {\em ad hoc} lattice region,
spatially separated from the one where computation has to take
place, to be subsequently loaded into the latter via the transport
mechanism detailed above.

\subsection{Single-qubit gates}
In the single-qubit case, the relevant resonance field is
$B_{0x}$, while the lasers couple $|1x\rangle$ with the lower
dressed state connected to $|0x\rangle$ with an effective Rabi
frequency $\Omega_1$ as in Fig.~\ref{doublevels}.b. The process is
resonant only if the marker atom in state $|x\rangle$ is present.
In this way, specific sites where the single-qubit operation takes
place can be selected even if the addressing laser cannot resolve
them spatially from neighboring sites. Moreover, the two-atoms
state remains always factorized, whence possible magnetic field
fluctuations, affecting the state $|x\rangle$ (unlike $|0\rangle$
and $|1\rangle$) will yield only a global phase.
In a rotating frame the Hamiltonian which describes this system takes the form
\begin{align}
  H = & \delta \ket{1x}\bra{1x}
  + \varepsilon_{0x}(B)\ket{M}\bra{M} \nonumber\\
  +&V_0^{0x}\big( \ket{0x}\bra{M}+\ket{M}\bra{0x} \big)
  + \frac{\Omega_1}{2}\big( \ket{0x}\bra{1x}+\ket{1x}\bra{0x} \big) \label{single}
\end{align}
where $\delta$ is the Raman detuning of two co-propagating Raman lasers, $V_0^{0x}$ is the coupling
between the molecular state $\ket{M}$ and the dissociated state
$\ket{0x}$ and $\varepsilon_{0x}(B)$ is the energy of the molecular
state (we set the energy of the $\ket{0x}$ state to zero).  At the
beginning of the operation the lasers are switched off (i.e.
$\Omega_1=0$) and the external magnetic field is adiabatically
tuned to the Feshbach resonance, i.e.
$\varepsilon_{0x}(B)\rightarrow\varepsilon_{0x}(B_{0x}) =0$. This leads to a
splitting of the two particle state $\ket{0x}$ in two new
eigenstates,
\begin{equation}
\ket{S_\pm^{0x}} = \frac{1}{\sqrt 2}\big( \ket{0x} \pm \ket{M} \big). \\
\end{equation}
These states are indicated in the left hand side of
Fig.~\ref{singlevels} (here we have $\Delta\epsilon^{0x}=V^{0x}_0$
since we are on resonance).  For finite laser power, in this basis the
Hamiltonian takes the form
\begin{align}
  H = &\delta\ket{1x}\bra{1x}
  + V_0^{0x} (\ket{S_+^{0x}}\bra{S_+^{0x}} - \ket{S_-^{0x}}\bra{S_-^{0x}}) \nonumber\\
  &+ \frac{\Omega_R}{2\sqrt 2}\Big[
  \big(\ket{S_+^{0x}}+\ket{S_-^{0x}}\big)\bra{1x}+\ket{1x}\big(\bra{S_+^{0x}}
  + \bra{S_-^{0x}}\big) \Big].
\end{align}
The Raman detuning is set to $\delta = -V_0^{0x}$ and if
$\Omega_0/V_0^{0x}\ll 1$ we can project out the state
$\ket{S_+^{0x}}$. This yields the two level Hamiltonian
\begin{align}
  H = & -V_0^{0x}\big( \ket{1x}\bra{1x} + \ket{S_-^{0x}}\bra{S_-^{0x}} \big) \nonumber\\
  & + \frac{\Omega_1}{2\sqrt
    2}\Big(\ket{S_-^{0x}}\bra{1x}+\ket{1x}\bra{S_-^{0x}} \Big)
\end{align}
i.e. if we finally tune the magnetic field out of the Feshbach
resonance again we get the transformation
\begin{align}
  \ket{0x}\rightarrow &\cos(\Omega_1/2\sqrt2t)\ket{1x} -\ii\sin(\Omega_1/2\sqrt2t)\e^{\ii\varphi} \ket{0x}, \\
  \ket{1x}\rightarrow &-\ii\sin(\Omega_1/2\sqrt2t) \ket{1x} +
  \cos(\Omega_1/2\sqrt2t)\e^{\ii\varphi}\ket{0x}.
\end{align}
In this expression we included a phase $\varphi$ which is the
(adjustable) phase accumulated during the adiabatic ramping process of
the magnetic field.
\subsection{Two-qubit gates}
In the two-qubit case, we take the marker atom to be in a state of the
logical subspace spanned by $|0\rangle$ and $|1\rangle$.  This time,
the field is ramped across $B_{00}$, and the Raman lasers couple for a
time $\tau$ --~with Rabi frequency $\Omega_2$~-- the lower dressed
state to the degenerate two-atom levels $|01\rangle$ and $|10\rangle$
(Fig.~\ref{doublevels}.c).  In a rotating frame the Hamiltonian can be
written as
\begin{align}
  H = &\delta \big(\ket{01}\bra{01} + \ket{10}\bra{10}\big) + 2\delta\ket{11}\bra{11} \nonumber\\
  &+ \varepsilon_{00}(B)\ket{M}\bra{M} + V_0^{00}\big(\ket{M}\bra{00} + \ket{00}\bra{M} \big) \nonumber\\
  &+ \frac{\Omega_2}{2}\big( \ket{00}\bra{01} + \ket{00}\bra{10} +
  \ket{11}\bra{01} +\ket{11}\bra{10} + \text{h.c.} \big).
\end{align}
The notations are the same as in Eq.~(\ref{single}). We perform now
the same procedure as in the case of the single qubit rotation, i.e.
we tune adiabatically the magnetic field to the Feshbach resonance,
i.e. $\varepsilon_{00}(B_{00}) = 0$ while $\Omega_2=0$. The Hamiltonian
with diagonalized molecular part reads
\begin{align}
  H = &\delta\big(\ket{01}\bra{01} + \ket{10}\bra{10}\big) + 2\delta\ket{11}\bra{11} \nonumber\\
  &+ V_0^{00}\big(\ket{S_+^{00}}\bra{S_+^{00}} - \ket{S_-^{00}}\bra{S_-^{00}}\big) \nonumber\\
  &+ \frac{\Omega_R}{2\sqrt 2}\Big[ \big(\ket{S_+^{00}} + \ket{S_-^{00}}\big)\bra{01} +  \big(\ket{S_+^{00}} + \ket{S_-^{00}}\big) \bra{10} \nonumber\\
  &+ \ket{11}\bra{01} +\ket{11}\bra{10} + \text{h.c.} \Big],
\end{align}
where
\begin{equation}
\ket{S_\pm^{00}} = \frac{1}{\sqrt 2}\big( \ket{00} \pm \ket{M} \big),\\
\end{equation}
These states are shown on the right hand side of
Fig.~\ref{singlevels} (now we have
$\Delta\epsilon^{00}=V^{00}_0$).  Taking the Raman detuning to be
$\delta = -V_0^{00}$ amounts to the fact that (if
$\Omega_2/V_0^{00}\ll1$) the states $\ket{11}$ and
$\ket{S_+^{00}}$ are effectively decoupled from the remaining
three states. Projecting out the uncoupled states the effective
Hamiltonian for the remaining three level system then takes the
form
\begin{align}
  H =& -V_0^{00}\big( \ket{S_-^{00}}\bra{S_-^{00}} + \ket{01}\bra{01} +\ket{10}\bra{10} \big) \nonumber\\
  & + \frac{\Omega_0}{2\sqrt 2}\big( \ket{S_-^{00}}\bra{01} +
  \ket{S_-^{00}}\bra{10} + \ket{01}\bra{S_-^{00}} +
  \ket{10}\bra{S_-^{00}} \big).
\end{align}
If we introduce the vector notation $\ket{\psi} \leftrightarrow
\big(\langle S_-^{00}\vert \psi\rangle,\,\langle 01\vert
\psi\rangle,\,\langle 10\vert \psi\rangle\big)^T$ and disregard global
phases the time evolution operator of this system can be written as
\begin{align}
  U(t) =
  &\frac{1}{2}\begin{pmatrix}
    2c(t) & -\ii s(t) & -\ii s(t) \\
    -\ii s(t)  & c(t)+1 & c(t)-1 \\
    -\ii s(t) & c(t)-1 & c(t)+1
\end{pmatrix}
\end{align}
with $c(t) = \cos(\Omega_2t/2)$ and $s(t) = \cos(\Omega_2t/2)$.  If we
apply a Raman pulse of duration $\tau=2(2n+1)\pi/\Omega_2$ and finally
tune the magnetic field out of the Feshbach resonance again, we get
the following truth table for the operation:
\begin{align}
  & \ket{00}\longrightarrow -\e^{\ii\varphi}\ket{00}, \nonumber\\
  & \ket{01}\longrightarrow -\ket{01},  \nonumber\\
  & \ket{10}\longrightarrow -\ket{10}, \nonumber\\
  & \ket{11}\longrightarrow \e^{\ii V_0^{00}\tau}\ket{11},
\end{align}
where we included again the phase $\varphi$ now accumulated by
state $|00\rangle$ during the ramping process due to the
interaction energy shift, whose value can be adjusted by
controlling the magnetic field. For $\varphi=2\pi$ and
$V_0^{00}\tau=(2m+1)\pi$ --~which imposes a commensurability
condition $\Omega_2/V_0^{00}=2(2n+1)/(2m+1)$ between the Rabi
frequency and the Feshbach energy shift~--, a swap operation is
performed. Besides being an essential ingredient for entangling
gates between distant atoms as detailed in Sec.~\ref{sec:sec2},
such a swap operation can greatly help in the task of
non-destructive qubit readout. To this aim, the quantum state of
an atom to be read out at the end of a computation could be simply
swapped onto a marker atom to be subsequently transported to a
different lattice region where measurement can take place without
physically disturbing the register atoms, which can be later
re-used for logical operations.
\begin{figure}[t]
  \begin{center}
    \includegraphics[width=7truecm]{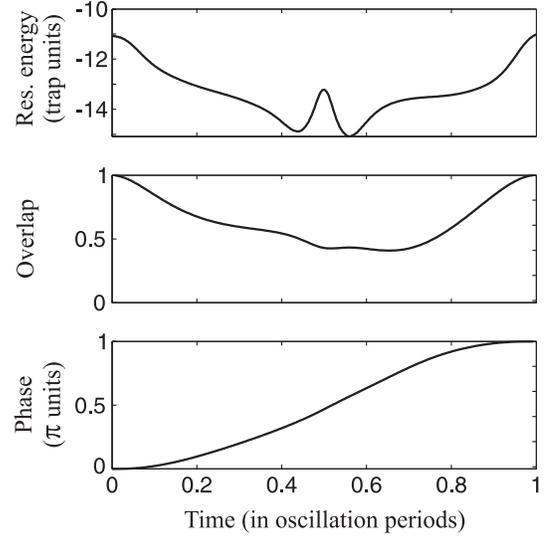}
    \caption{
      Two-qubit (C-phase) gate operation: resonance energy
      $\varepsilon_{00}$ (top; off-resonant before and after gate
      operation), overlap between evolved and initial state (center),
      accumulated phase (bottom).}
    \label{gate}
  \end{center}
\end{figure}

On the other hand, if no Raman lasers are present and $\varphi=\pi$, a
C-phase gate between register and marker atom is obtained. A two qubit
gate between distant register atoms can be realized as described in
Sec.~\ref{sec:sec2}. Note that laser addressing of single qubits is
never required throughout the procedure.

The magnetic ramping process can be even performed non-adiabatically,
provided that all population is finally returned to the trapped atomic
ground state. This can be accomplished via a quantum optimal control
technique in analogy with the above discussion for the transport
process. The control parameter in this case is the resonance energy
$\varepsilon_{00}$, which can be adjusted by varying the external
magnetic field. Care has to be taken in optimizing not only the
absolute value of the overlap of the final state onto the goal state,
but also its phase $\varphi$. Fig.~\ref{gate} shows the optimization
results for a 100 kHz trap with a ratio of $\nu_\perp/\nu=10$ between the trap frequencies $\nu_\perp$ in $y,z$-direction and $\nu$ in $x$-direction. The
final infidelity is about $2\times10^{-5}$ in this case.

\section{\label{concl}Conclusions}
When it comes to using neutral atoms for the purpose of quantum
information processing, besides the well-known general criteria
formulated by D. DiVincenzo \cite{DiVincenzoFdP}, the fulfillment
of various practical requirements, specific to atomic
implementations, can make a difference on the road to experimental
realization. For example laser addressing of single qubits, though being
theoretically trivial, is limited by diffraction, imposing a lower
bound on the actual spacing between qubits. Furthermore performing gate operations in state-dependent potentials creates entanglement between
internal and external degrees of freedom, which in turn is prone to
decoherence, as random fields typically affect differently the two
logical states. The same is true for internal-state
entanglement, if the qubit states are chosen with different Land\'e
factor and, unless the latter vanishes for both states, they will be
sensitive to magnetic-field fluctuations.

In this paper, we introduced the concept of ``marking'' qubits
via molecular interactions which allows for relaxing a number of
these constraints for neutral-atom quantum computing. We have presented
a scheme that enables quantum gates and information transport in a
quantum register, even though requiring neither single-site
addressing by externally applied fields nor state-dependent
external potential. Moreover, qubit states with the same (even
vanishing) Land\'e factor can be employed; and the overall speed
can be of the order of the inverse atomic trapping frequency.
We have shown how this scheme can be implemented in two-component
optical lattices, whereby the mechanism used to mark atoms is the
molecular interaction responsible for Feshbach resonances, which
are currently a subject of intense experimental research in the
field of cold atoms, where molecule formation via control of
Feshbach resonances has been recently achieved
\cite{ketterle03,wieman02,regal03,grimm03,salomon03,rempe04}. In
other words, our proposal relies on techniques that are presently
being developed, and represents therefore a feasible candidate for
the implementation of quantum information processing with neutral
atoms in optical lattices.

Finally, the analysis presented here is limited to one-dimensional
systems, basically with a single marker atom. Further conceptual
development is possible, for instance in exploring the interplay
between several marker atoms on the same lattice, or the extended
flexibility given for instance by higher-dimensional geometries;
this will be the subject of future investigations.

\acknowledgments We gratefully acknowledge inspiring discussions
with E. Tiesinga and S. Sklarz. This work has been co-financed by
MIUR and supported by a Fulbright grant, the Austrian Science
Foundation FWF, the European Commission under contracts
IST-2001-38863 (ACQP) and HPRN-CT-2000-00121 (QUEST), and the
Institute for Quantum Information. T.C. thanks NIST Gaithersburg
for its warm hospitality.

\appendix

\section{Dynamics in a cigar-shaped trap\label{app:cigar}}

The normalized eigenfunctions of a 3D harmonic oscillator in spherical coordinates are%
\begin{eqnarray}
\Phi _{n\ell m_{\ell }}^{\mathrm{sph}}(r,\theta ,\phi
)&=&\sqrt{\frac{2\alpha ^{2\ell +3}\Gamma (n+\ell +3/2)}{n!\Gamma
(\ell +3/2)^{2}}}r^{\ell }e^{-\alpha ^{2}r^{2}/2}\times\nonumber\\
&&_{1}F_{1}(-n,\ell +3/2,\alpha ^{2}r^{2})Y_{\ell }^{m_{\ell
}}(\theta ,\phi )
\end{eqnarray}%
where $\alpha =\sqrt{m\nu /\hbar }$, $\Gamma (n)$ is the Euler
Gamma function, $_{1}F_{1}(a,b,z)$ is the Kummer confluent
hypergeometric function, and $Y_{\ell }^{m_{\ell }}(\theta ,\phi
)$ are the spherical harmonics. The normalized eigenfunctions in
cylindrical coordinates (we assume the same frequency $\nu $ in
the longitudinal direction, and a
transverse frequency $\nu_\perp$ a factor $\gamma $ higher) are:%
\begin{eqnarray}
\Phi _{n\ell m_{\ell }}^{\mathrm{cyl}}(\rho ,\phi
,z)&=&\frac{(\alpha \rho )^{|m_{\ell }|}e^{im_{\ell }\phi -(\alpha
^{2}\gamma \rho ^{2}+\alpha
^{2}z^{2})/2}}{\sqrt{2^{n}\ell !(\sqrt{\pi }%
/\alpha )^{3}}}\times\\
&&H_{n}(\alpha z)\,_{1}F_{1}(-\ell ,\left\vert m_{\ell
}\right\vert +1,\gamma \alpha ^{2}\rho ^{2})/\nonumber\\
&&\sqrt{\sum_{i,j=0}^{\ell }\frac{(-\ell )_{i}(-\ell )_{j}\Gamma
(|m_{\ell }|+i+j+1)}{i!j!(m_{\ell }+1)_{i}(m_{\ell }+1)_{j}\gamma
^{|m_{\ell }|+1}}}\nonumber
\end{eqnarray}%
where $H_{n}(x)$ are the Hermite polynomials, and $(a)_{i}=\Gamma
(a+i)/\Gamma (a)$ is the Pochhammer symbol. We are interested in
$s$-wave
scattering processes, so we restrict our analysis to the eigenstates with $%
\ell =m_{\ell }=0$ and obtain
\begin{eqnarray}
\langle \Phi _{2v+1,0,0}^{\mathrm{cyl}}|\Phi
_{w,0,0}^{\mathrm{sph}}\rangle
\!&\!=\!&\!0; \\
\langle \Phi _{2v,0,0}^{\mathrm{cyl}}|\Phi
_{w,0,0}^{\mathrm{sph}}\rangle \!&\!=\!&\!(-2)^{-v}\sqrt{\gamma
2^{3-w}(2v)!w!(2w+1)!!}\nonumber\\
\!&\!\times\!&\!\sum_{i=0}^{w}\sum_{j=0}^{v}\frac{(-4)^{i+j}(2i+2j+1)!!}
{(2i+1)!(2j+1)!}\qquad\\
\!&\!\times\!&\!\frac{_{1}F_{1}\left(
j+\frac{1}{2},i+j+\frac{3}{2},j+\frac{3}{2},\frac{\gamma -1}{\gamma +1}%
\right)}{\left( v-j\right) !(w-i)!(\gamma +1)^{i+j+3/2}}\nonumber
\end{eqnarray}%

The coupling matrix elements in a cigar-shaped trap with anisotropy factor $%
\gamma \neq 1$ are computed as
\begin{equation}
V_{v}^\beta(\gamma )=\sum_{w}\langle \Phi _{v00}^{\mathrm{cyl}}|\Phi _{w00}^{%
\mathrm{sph}}\rangle V_{w}^\beta(\gamma =1),
\end{equation}%
where the spherical matrix elements $V_{w}(\gamma =1)$ are given by Eq. (\ref%
{couplings}).

\section{Conditional level shift in a quasi-1D trap\label{app:levels}}

Let us consider the three-dimensional state $\left\vert \Psi
_{S,m_{S}}\right\rangle $ of two spin-1/2 bosons in a harmonic trap. In the $%
\hat{x}$ direction, one particle is in the trap ground state
$\left\vert
\psi _{0}\right\rangle $, and the other in the first excited state $%
\left\vert \psi _{1}\right\rangle $. The transverse state is the
ground
state $\left\vert \psi _{\perp }\right\rangle $ for both particles. $X$ and $%
x$ are the center-of-mass and relative coordinate. Denoting the
$i^{\rm th}$ particle's state by $|\cdot\rangle_i$, the
symmetrized states can be written as
\begin{align}
\left\vert \Psi _{0,0}\right\rangle
&=\frac{\left\vert 1\right\rangle _{1}\left\vert 0 \right\rangle _{2}-\left\vert
0 \right\rangle _{1}\left\vert 1 \right\rangle _{2}}{\sqrt{2}%
}\frac{|\psi_{1}\rangle_1|\psi_{0}\rangle_2-|\psi_{0}\rangle_1|\psi_{1}\rangle_2}{%
\sqrt{2}}\nonumber\\
&=\frac{\left\vert 1 \right\rangle _{1}\left\vert 0 \right\rangle
_{2}-\left\vert 0 \right\rangle _{1}\left\vert 1 \right\rangle
_{2}}{\sqrt{2}}|\psi_{0}\rangle_X|\psi_{1}\rangle_x; \\
\left\vert \Psi _{1,-1}\right\rangle
&=\left\vert 0\right\rangle _{1}\left\vert 0 \right\rangle
_{2}\frac{|\psi_{1}\rangle_1|\psi_{0}\rangle_2+|\psi_{0}\rangle_1|\psi_{1}\rangle_2}{\sqrt{2}}
\nonumber\\
&=\left\vert 0 \right\rangle _{1}\left\vert 0 \right\rangle
_{2}|\psi_{1}\rangle_X|\psi_{0}\rangle_x;
\end{align}
\begin{align}
\left\vert \Psi _{1,0}\right\rangle
&=\frac{\left\vert 1\right\rangle _{1}\left\vert 0 \right\rangle _{2}+\left\vert
0 \right\rangle _{1}\left\vert 1 \right\rangle _{2}}{\sqrt{2}%
}\frac{|\psi_{1}\rangle_1|\psi_{0}\rangle_2+|\psi_{0}\rangle_1|\psi_{1}\rangle_2}{%
\sqrt{2}} \nonumber\\
&=\frac{\left\vert 1 \right\rangle _{1}\left\vert 0 \right\rangle
_{2}+\left\vert 0 \right\rangle _{1}\left\vert 1 \right\rangle
_{2}}{\sqrt{2}}|\psi_{1}\rangle_X|\psi_{0}\rangle_x; \\
\left\vert \Psi _{1,1}\right\rangle
&=\left\vert 1 \right\rangle_{1}\left\vert 1 \right\rangle
_{2}\frac{|\psi_{1}\rangle_1|\psi_{0}\rangle_2+|\psi_{0}\rangle_1|\psi_{1}\rangle_2}{\sqrt{2}}
\nonumber \\
&=\left\vert 1 \right\rangle _{1}\left\vert 1 \right\rangle
_{2}|\psi_{1}\rangle_X|\psi_{0}\rangle_x.
\end{align}%
When we apply a static external magnetic field corresponding to the
Feshbach resonance for the $|00\rangle$ channel,
the interaction only affects the state $%
\left\vert \Psi _{1,-1}\right\rangle $, dressing it with a splitting $%
2V_{0}^{00}(\gamma)$ that can easily be of the order of $\hbar\nu$
assuming a ratio $\nu_\perp/\nu=10$ of the trap frequancy $\nu_\perp$
in $y,z$-direction and $\nu$ in $x$-direction. This means that the
state $\left\vert 0 \right\rangle _{1}\left\vert 0 \right\rangle _{2}$
can be discriminated spectroscopically, allowing for different kinds
of gate operation as described in the text.


\begin{thebibliography}{41}
\expandafter\ifx\csname
natexlab\endcsname\relax\def\natexlab#1{#1}\fi
\expandafter\ifx\csname bibnamefont\endcsname\relax
  \def\bibnamefont#1{#1}\fi
\expandafter\ifx\csname bibfnamefont\endcsname\relax
  \def\bibfnamefont#1{#1}\fi
\expandafter\ifx\csname citenamefont\endcsname\relax
  \def\citenamefont#1{#1}\fi
\expandafter\ifx\csname url\endcsname\relax
  \def\url#1{\texttt{#1}}\fi
\expandafter\ifx\csname
urlprefix\endcsname\relax\def\urlprefix{URL }\fi
\providecommand{\bibinfo}[2]{#2}
\providecommand{\eprint}[2][]{\url{#2}}

\bibitem[{\citenamefont{Cirac and Zoller}(2004)}]{cirac04}
\bibinfo{author}{\bibfnamefont{J.~I.} \bibnamefont{Cirac}} \bibnamefont{and}
  \bibinfo{author}{\bibfnamefont{P.}~\bibnamefont{Zoller}},
  \bibinfo{journal}{Phys. Today} \textbf{\bibinfo{volume}{57}},
  \bibinfo{pages}{38} (\bibinfo{year}{2004}).

\bibitem[{\citenamefont{Greiner et~al.}(2002)\citenamefont{Greiner, Mandel,
  Esslinger, H{\"a}nsch, and Bloch}}]{SFMott}
\bibinfo{author}{\bibfnamefont{M.}~\bibnamefont{Greiner}},
  \bibinfo{author}{\bibfnamefont{O.}~\bibnamefont{Mandel}},
  \bibinfo{author}{\bibfnamefont{T.}~\bibnamefont{Esslinger}},
  \bibinfo{author}{\bibfnamefont{T.~W.} \bibnamefont{H{\"a}nsch}},
  \bibnamefont{and} \bibinfo{author}{\bibfnamefont{I.}~\bibnamefont{Bloch}},
  \bibinfo{journal}{Nature} \textbf{\bibinfo{volume}{415}}, \bibinfo{pages}{39}
  (\bibinfo{year}{2002}).

\bibitem[{\citenamefont{Charron et~al.}(2002)\citenamefont{Charron, Tiesinga,
  Mies, and Williams}}]{charron02}
\bibinfo{author}{\bibfnamefont{E.}~\bibnamefont{Charron}},
  \bibinfo{author}{\bibfnamefont{E.}~\bibnamefont{Tiesinga}},
  \bibinfo{author}{\bibfnamefont{F.}~\bibnamefont{Mies}}, \bibnamefont{and}
  \bibinfo{author}{\bibfnamefont{C.}~\bibnamefont{Williams}},
  \bibinfo{journal}{Phys. Rev. Lett.} \textbf{\bibinfo{volume}{88}},
  \bibinfo{pages}{077901} (\bibinfo{year}{2002}).

\bibitem[{\citenamefont{Tian and Zoller}(2003)}]{tian03}
\bibinfo{author}{\bibfnamefont{L.}~\bibnamefont{Tian}} \bibnamefont{and}
  \bibinfo{author}{\bibfnamefont{P.}~\bibnamefont{Zoller}},
  \bibinfo{journal}{Phys. Rev. A} \textbf{\bibinfo{volume}{68}},
  \bibinfo{pages}{042321} (\bibinfo{year}{2003}).

\bibitem[{\citenamefont{Pachos and Knight}(2003)}]{pachos03}
\bibinfo{author}{\bibfnamefont{J.~K.} \bibnamefont{Pachos}} \bibnamefont{and}
  \bibinfo{author}{\bibfnamefont{P.~L.} \bibnamefont{Knight}},
  \bibinfo{journal}{Phys. Rev. Lett.} \textbf{\bibinfo{volume}{91}},
  \bibinfo{pages}{107902} (\bibinfo{year}{2003}).

\bibitem[{\citenamefont{Dorner et~al.}(2003)\citenamefont{Dorner, Fedichev,
  Jaksch, and Zoller}}]{dorner03}
\bibinfo{author}{\bibfnamefont{U.}~\bibnamefont{Dorner}},
  \bibinfo{author}{\bibfnamefont{P.}~\bibnamefont{Fedichev}},
  \bibinfo{author}{\bibfnamefont{D.}~\bibnamefont{Jaksch}}, \bibnamefont{and}
  \bibinfo{author}{\bibfnamefont{P.}~\bibnamefont{Zoller}},
  \bibinfo{journal}{Phys. Rev. Lett.} \textbf{\bibinfo{volume}{91}},
  \bibinfo{pages}{073601} (\bibinfo{year}{2003}).

\bibitem[{\citenamefont{Rabl et~al.}(2003)\citenamefont{Rabl, Daley, Fedichev,
  Cirac, and Zoller}}]{rabl03}
\bibinfo{author}{\bibfnamefont{P.}~\bibnamefont{Rabl}},
  \bibinfo{author}{\bibfnamefont{A.~J.} \bibnamefont{Daley}},
  \bibinfo{author}{\bibfnamefont{P.~O.} \bibnamefont{Fedichev}},
  \bibinfo{author}{\bibfnamefont{J.~I.} \bibnamefont{Cirac}}, \bibnamefont{and}
  \bibinfo{author}{\bibfnamefont{P.}~\bibnamefont{Zoller}},
  \bibinfo{journal}{Phys. Rev. Lett.} \textbf{\bibinfo{volume}{91}},
  \bibinfo{pages}{110403} (\bibinfo{year}{2003}).

\bibitem[{\citenamefont{Mompart et~al.}(2003)\citenamefont{Mompart, Eckert,
  Ertmer, Birkl, and Lewenstein}}]{mompart03}
\bibinfo{author}{\bibfnamefont{J.}~\bibnamefont{Mompart}},
  \bibinfo{author}{\bibfnamefont{K.}~\bibnamefont{Eckert}},
  \bibinfo{author}{\bibfnamefont{W.}~\bibnamefont{Ertmer}},
  \bibinfo{author}{\bibfnamefont{G.}~\bibnamefont{Birkl}}, \bibnamefont{and}
  \bibinfo{author}{\bibfnamefont{M.}~\bibnamefont{Lewenstein}},
  \bibinfo{journal}{Phys. Rev. Lett.} \textbf{\bibinfo{volume}{90}},
  \bibinfo{pages}{147901} (\bibinfo{year}{2003}).

\bibitem[{\citenamefont{Eckert et~al.}(2002)\citenamefont{Eckert, Mompart, Yi,
  Schliemann, Bru{\ss}, Birkl, and Lewenstein}}]{eckert02}
\bibinfo{author}{\bibfnamefont{K.}~\bibnamefont{Eckert}},
  \bibinfo{author}{\bibfnamefont{J.}~\bibnamefont{Mompart}},
  \bibinfo{author}{\bibfnamefont{X.~X.} \bibnamefont{Yi}},
  \bibinfo{author}{\bibfnamefont{J.}~\bibnamefont{Schliemann}},
  \bibinfo{author}{\bibfnamefont{D.}~\bibnamefont{Bru{\ss}}},
  \bibinfo{author}{\bibfnamefont{G.}~\bibnamefont{Birkl}}, \bibnamefont{and}
  \bibinfo{author}{\bibfnamefont{M.}~\bibnamefont{Lewenstein}},
  \bibinfo{journal}{Phys. Rev. A} \textbf{\bibinfo{volume}{66}},
  \bibinfo{pages}{042317} (\bibinfo{year}{2002}).

\bibitem[{\citenamefont{Lukin et~al.}(2001)\citenamefont{Lukin, Fleischhauer,
  C{\^o}t{\'e}, Duan, Jaksch, Cirac, and Zoller}}]{lukin01}
\bibinfo{author}{\bibfnamefont{M.~D.} \bibnamefont{Lukin}},
  \bibinfo{author}{\bibfnamefont{M.}~\bibnamefont{Fleischhauer}},
  \bibinfo{author}{\bibfnamefont{R.}~\bibnamefont{C{\^o}t{\'e}}},
  \bibinfo{author}{\bibfnamefont{L.~M.} \bibnamefont{Duan}},
  \bibinfo{author}{\bibfnamefont{D.}~\bibnamefont{Jaksch}},
  \bibinfo{author}{\bibfnamefont{J.~I.} \bibnamefont{Cirac}}, \bibnamefont{and}
  \bibinfo{author}{\bibfnamefont{P.}~\bibnamefont{Zoller}},
  \bibinfo{journal}{Phys. Rev. Lett.} \textbf{\bibinfo{volume}{87}},
  \bibinfo{pages}{037901} (\bibinfo{year}{2001}).

\bibitem[{\citenamefont{Andersson and Stenholm}(2001)}]{andersson01}
\bibinfo{author}{\bibfnamefont{E.}~\bibnamefont{Andersson}} \bibnamefont{and}
  \bibinfo{author}{\bibfnamefont{S.}~\bibnamefont{Stenholm}},
  \bibinfo{journal}{Opt. Commun.} \textbf{\bibinfo{volume}{188}},
  \bibinfo{pages}{141} (\bibinfo{year}{2001}).

\bibitem[{\citenamefont{Garc{\'i}a-Ripoll and Cirac}(2003)}]{ripoll03}
\bibinfo{author}{\bibfnamefont{J.~J.} \bibnamefont{Garc{\'i}a-Ripoll}}
  \bibnamefont{and} \bibinfo{author}{\bibfnamefont{J.~I.} \bibnamefont{Cirac}},
  \bibinfo{journal}{Phys. Rev. Lett.} \textbf{\bibinfo{volume}{90}},
  \bibinfo{pages}{127902} (\bibinfo{year}{2003}).

\bibitem[{\citenamefont{Brennen et~al.}(2000)\citenamefont{Brennen, Deutsch,
  and Jessen}}]{brennen00}
\bibinfo{author}{\bibfnamefont{G.~K.} \bibnamefont{Brennen}},
  \bibinfo{author}{\bibfnamefont{I.~H.} \bibnamefont{Deutsch}},
  \bibnamefont{and} \bibinfo{author}{\bibfnamefont{P.~S.}
  \bibnamefont{Jessen}}, \bibinfo{journal}{Phys. Rev. A}
  \textbf{\bibinfo{volume}{61}}, \bibinfo{pages}{062309}
  (\bibinfo{year}{2000}).

\bibitem[{\citenamefont{Brennen et~al.}(1999)\citenamefont{Brennen, Caves,
  Jessen, and Deutsch}}]{brennen99}
\bibinfo{author}{\bibfnamefont{G.~K.} \bibnamefont{Brennen}},
  \bibinfo{author}{\bibfnamefont{C.~M.} \bibnamefont{Caves}},
  \bibinfo{author}{\bibfnamefont{P.~S.} \bibnamefont{Jessen}},
  \bibnamefont{and} \bibinfo{author}{\bibfnamefont{I.~H.}
  \bibnamefont{Deutsch}}, \bibinfo{journal}{Phys. Rev. Lett.}
  \textbf{\bibinfo{volume}{82}}, \bibinfo{pages}{1060} (\bibinfo{year}{1999}).

\bibitem[{\citenamefont{Schlosser et~al.}(2001)\citenamefont{Schlosser,
  Reymond, Protsenko, and Grangier}}]{grangier01}
\bibinfo{author}{\bibfnamefont{N.}~\bibnamefont{Schlosser}},
  \bibinfo{author}{\bibfnamefont{G.}~\bibnamefont{Reymond}},
  \bibinfo{author}{\bibfnamefont{I.}~\bibnamefont{Protsenko}},
  \bibnamefont{and} \bibinfo{author}{\bibfnamefont{P.}~\bibnamefont{Grangier}},
  \bibinfo{journal}{Nature} \textbf{\bibinfo{volume}{411}},
  \bibinfo{pages}{1024} (\bibinfo{year}{2001}).

\bibitem[{\citenamefont{Dumke et~al.}(2002)\citenamefont{Dumke, M{\"u}ther,
  Volk, Ertmer, and Birkl}}]{dumke02}
\bibinfo{author}{\bibfnamefont{R.}~\bibnamefont{Dumke}},
  \bibinfo{author}{\bibfnamefont{T.}~\bibnamefont{M{\"u}ther}},
  \bibinfo{author}{\bibfnamefont{M.}~\bibnamefont{Volk}},
  \bibinfo{author}{\bibfnamefont{W.}~\bibnamefont{Ertmer}}, \bibnamefont{and}
  \bibinfo{author}{\bibfnamefont{G.}~\bibnamefont{Birkl}},
  \bibinfo{journal}{Phys. Rev. Lett.} \textbf{\bibinfo{volume}{89}},
  \bibinfo{pages}{220402} (\bibinfo{year}{2002}).

\bibitem[{\citenamefont{D{\"u}r et~al.}(1999)\citenamefont{D{\"u}r, Briegel,
  Cirac, and Zoller}}]{repeater}
\bibinfo{author}{\bibfnamefont{W.}~\bibnamefont{D{\"u}r}},
  \bibinfo{author}{\bibfnamefont{H.}~\bibnamefont{Briegel}},
  \bibinfo{author}{\bibfnamefont{J.~I.} \bibnamefont{Cirac}}, \bibnamefont{and}
  \bibinfo{author}{\bibfnamefont{P.}~\bibnamefont{Zoller}},
  \bibinfo{journal}{Phys. Rev. A} \textbf{\bibinfo{volume}{59}},
  \bibinfo{pages}{169} (\bibinfo{year}{1999}).

\bibitem[{\citenamefont{Folman et~al.}(2002)\citenamefont{Folman, Krueger,
  Schmiedmayer, Denschlag, and Henkel}}]{folman02}
\bibinfo{author}{\bibfnamefont{R.}~\bibnamefont{Folman}},
  \bibinfo{author}{\bibfnamefont{P.}~\bibnamefont{Krueger}},
  \bibinfo{author}{\bibfnamefont{J.}~\bibnamefont{Schmiedmayer}},
  \bibinfo{author}{\bibfnamefont{J.}~\bibnamefont{Denschlag}},
  \bibnamefont{and} \bibinfo{author}{\bibfnamefont{C.}~\bibnamefont{Henkel}},
  \bibinfo{journal}{Adv. At. Mol. Opt. Phys.} \textbf{\bibinfo{volume}{48}},
  \bibinfo{pages}{263} (\bibinfo{year}{2002}).

\bibitem[{\citenamefont{Farooqi et~al.}(2003)\citenamefont{Farooqi, Tong,
  Krishnan, Stanojevic, Zhang, Ensher, Estrin, Boisseau, C{\^o}t{\'e}, Eyler
  et~al.}}]{cote03}
\bibinfo{author}{\bibfnamefont{S.~M.} \bibnamefont{Farooqi}},
  \bibinfo{author}{\bibfnamefont{D.}~\bibnamefont{Tong}},
  \bibinfo{author}{\bibfnamefont{S.}~\bibnamefont{Krishnan}},
  \bibinfo{author}{\bibfnamefont{J.}~\bibnamefont{Stanojevic}},
  \bibinfo{author}{\bibfnamefont{Y.~P.} \bibnamefont{Zhang}},
  \bibinfo{author}{\bibfnamefont{J.~R.} \bibnamefont{Ensher}},
  \bibinfo{author}{\bibfnamefont{A.~S.} \bibnamefont{Estrin}},
  \bibinfo{author}{\bibfnamefont{C.}~\bibnamefont{Boisseau}},
  \bibinfo{author}{\bibfnamefont{R.}~\bibnamefont{C{\^o}t{\'e}}},
  \bibinfo{author}{\bibfnamefont{E.~E.} \bibnamefont{Eyler}},
  \bibnamefont{et~al.}, \bibinfo{journal}{Phys. Rev. Lett.}
  \textbf{\bibinfo{volume}{91}}, \bibinfo{pages}{183002}
  (\bibinfo{year}{2003}).

\bibitem[{\citenamefont{Duan et~al.}(2004)\citenamefont{Duan, Blinov, Moehring,
  and Monroe}}]{duan04}
\bibinfo{author}{\bibfnamefont{L.-M.} \bibnamefont{Duan}},
  \bibinfo{author}{\bibfnamefont{B.~B.} \bibnamefont{Blinov}},
  \bibinfo{author}{\bibfnamefont{D.~L.} \bibnamefont{Moehring}},
  \bibnamefont{and} \bibinfo{author}{\bibfnamefont{C.}~\bibnamefont{Monroe}}
  (\bibinfo{year}{2004}), \eprint{quant-ph/0401020}.

\bibitem[{\citenamefont{Folman et~al.}(2000)\citenamefont{Folman, Kr{\"u}ger,
  Cassettari, Hessmo, Maier, and Schmiedmayer}}]{atomchip}
\bibinfo{author}{\bibfnamefont{R.}~\bibnamefont{Folman}},
  \bibinfo{author}{\bibfnamefont{P.}~\bibnamefont{Kr{\"u}ger}},
  \bibinfo{author}{\bibfnamefont{D.}~\bibnamefont{Cassettari}},
  \bibinfo{author}{\bibfnamefont{B.}~\bibnamefont{Hessmo}},
  \bibinfo{author}{\bibfnamefont{T.}~\bibnamefont{Maier}}, \bibnamefont{and}
  \bibinfo{author}{\bibfnamefont{J.}~\bibnamefont{Schmiedmayer}},
  \bibinfo{journal}{Phys. Rev. Lett.} \textbf{\bibinfo{volume}{84}},
  \bibinfo{pages}{4749} (\bibinfo{year}{2000}).

\bibitem[{\citenamefont{Pethick and Smith}(2002)}]{Feshbach}
\bibinfo{author}{\bibfnamefont{C.~J.} \bibnamefont{Pethick}} \bibnamefont{and}
  \bibinfo{author}{\bibfnamefont{H.}~\bibnamefont{Smith}},
  \emph{\bibinfo{title}{Bose-Einstein Condensation in Dilute Gases}}
  (\bibinfo{publisher}{Cambridge University Press},
  \bibinfo{address}{Cambridge}, \bibinfo{year}{2002}).

\bibitem[{\citenamefont{Jaksch et~al.}(1999)\citenamefont{Jaksch, Briegel,
  Cirac, Gardiner, and Zoller}}]{jaksch99}
\bibinfo{author}{\bibfnamefont{D.}~\bibnamefont{Jaksch}},
  \bibinfo{author}{\bibfnamefont{H.-J.} \bibnamefont{Briegel}},
  \bibinfo{author}{\bibfnamefont{J.~I.} \bibnamefont{Cirac}},
  \bibinfo{author}{\bibfnamefont{C.~W.} \bibnamefont{Gardiner}},
  \bibnamefont{and} \bibinfo{author}{\bibfnamefont{P.}~\bibnamefont{Zoller}},
  \bibinfo{journal}{Phys. Rev. Lett.} \textbf{\bibinfo{volume}{82}},
  \bibinfo{pages}{1975} (\bibinfo{year}{1999}).

\bibitem[{\citenamefont{Peil et~al.}(2003)\citenamefont{Peil, Porto, {Laburthe
  Tolra}, Obrecht, King, Subbotin, Rolston, and Phillips}}]{Rolston03}
\bibinfo{author}{\bibfnamefont{S.}~\bibnamefont{Peil}},
  \bibinfo{author}{\bibfnamefont{J.~V.} \bibnamefont{Porto}},
  \bibinfo{author}{\bibfnamefont{B.}~\bibnamefont{{Laburthe Tolra}}},
  \bibinfo{author}{\bibfnamefont{J.~M.} \bibnamefont{Obrecht}},
  \bibinfo{author}{\bibfnamefont{B.~E.} \bibnamefont{King}},
  \bibinfo{author}{\bibfnamefont{M.}~\bibnamefont{Subbotin}},
  \bibinfo{author}{\bibfnamefont{S.~L.} \bibnamefont{Rolston}},
  \bibnamefont{and} \bibinfo{author}{\bibfnamefont{W.~D.}
  \bibnamefont{Phillips}}, \bibinfo{journal}{Phys. Rev. A}
  \textbf{\bibinfo{volume}{67}}, \bibinfo{pages}{R051603}
  (\bibinfo{year}{2003}).

\bibitem[{\citenamefont{Press et~al.}(1992)\citenamefont{Press, Teukolsky,
  Vetterling, and Flannery}}]{NR}
\bibinfo{author}{\bibfnamefont{W.~H.} \bibnamefont{Press}},
  \bibinfo{author}{\bibfnamefont{S.~A.} \bibnamefont{Teukolsky}},
  \bibinfo{author}{\bibfnamefont{W.~T.} \bibnamefont{Vetterling}},
  \bibnamefont{and} \bibinfo{author}{\bibfnamefont{B.~P.}
  \bibnamefont{Flannery}}, \emph{\bibinfo{title}{Numerical Recipes in C}}
  (\bibinfo{publisher}{Cambridge University Press},
  \bibinfo{address}{Cambridge}, \bibinfo{year}{1992}), \bibinfo{edition}{2nd}
  ed.

\bibitem[{\citenamefont{Peirce et~al.}(1988)\citenamefont{Peirce, Dahleh, and
  Rabitz}}]{rabitz}
\bibinfo{author}{\bibfnamefont{A.~P.} \bibnamefont{Peirce}},
  \bibinfo{author}{\bibfnamefont{M.~A.} \bibnamefont{Dahleh}},
  \bibnamefont{and} \bibinfo{author}{\bibfnamefont{H.}~\bibnamefont{Rabitz}},
  \bibinfo{journal}{Phys. Rev. A} \textbf{\bibinfo{volume}{37}},
  \bibinfo{pages}{4950} (\bibinfo{year}{1988}).

\bibitem[{\citenamefont{Borzi et~al.}(2002)\citenamefont{Borzi, Stadler, and
  Hohenester}}]{hohenester}
\bibinfo{author}{\bibfnamefont{A.}~\bibnamefont{Borzi}},
  \bibinfo{author}{\bibfnamefont{G.}~\bibnamefont{Stadler}}, \bibnamefont{and}
  \bibinfo{author}{\bibfnamefont{U.}~\bibnamefont{Hohenester}},
  \bibinfo{journal}{Phys. Rev. A} \textbf{\bibinfo{volume}{66}},
  \bibinfo{pages}{053811} (\bibinfo{year}{2002}).

\bibitem[{\citenamefont{Sklarz and Tannor}(2002)}]{Sklarz02}
\bibinfo{author}{\bibfnamefont{S.}~\bibnamefont{Sklarz}} \bibnamefont{and}
  \bibinfo{author}{\bibfnamefont{D.}~\bibnamefont{Tannor}},
  \bibinfo{journal}{Phys. Rev. A} \textbf{\bibinfo{volume}{66}},
  \bibinfo{pages}{53619} (\bibinfo{year}{2002}).

\bibitem[{\citenamefont{Sola et~al.}(1998)\citenamefont{Sola, Santamaria, and
  Tannor}}]{sola}
\bibinfo{author}{\bibfnamefont{I.~R.} \bibnamefont{Sola}},
  \bibinfo{author}{\bibfnamefont{J.}~\bibnamefont{Santamaria}},
  \bibnamefont{and} \bibinfo{author}{\bibfnamefont{D.~J.}
  \bibnamefont{Tannor}}, \bibinfo{journal}{J. Phys. Chem.}
  \textbf{\bibinfo{volume}{102}}, \bibinfo{pages}{4301} (\bibinfo{year}{1998}).

\bibitem[{\citenamefont{Jessen and Deutsch}(1996)}]{optlatt}
\bibinfo{author}{\bibfnamefont{P.~S.} \bibnamefont{Jessen}} \bibnamefont{and}
  \bibinfo{author}{\bibfnamefont{I.~H.} \bibnamefont{Deutsch}},
  \bibinfo{journal}{Adv. At. Mol. Opt. Phys} \textbf{\bibinfo{volume}{37}},
  \bibinfo{pages}{95} (\bibinfo{year}{1996}).

\bibitem[{\citenamefont{Weiner et~al.}(1999)\citenamefont{Weiner, Bagnato,
  Zilio, and Julienne}}]{photoassociation}
\bibinfo{author}{\bibfnamefont{J.}~\bibnamefont{Weiner}},
  \bibinfo{author}{\bibfnamefont{V.~S.} \bibnamefont{Bagnato}},
  \bibinfo{author}{\bibfnamefont{S.}~\bibnamefont{Zilio}}, \bibnamefont{and}
  \bibinfo{author}{\bibfnamefont{P.~S.} \bibnamefont{Julienne}},
  \bibinfo{journal}{Rev. Mod. Phys.} \textbf{\bibinfo{volume}{71}},
  \bibinfo{pages}{1} (\bibinfo{year}{1999}).

\bibitem[{\citenamefont{Mies et~al.}(2000)\citenamefont{Mies, Tiesinga, and
  Julienne}}]{Mies00}
\bibinfo{author}{\bibfnamefont{F.~H.} \bibnamefont{Mies}},
  \bibinfo{author}{\bibfnamefont{E.}~\bibnamefont{Tiesinga}}, \bibnamefont{and}
  \bibinfo{author}{\bibfnamefont{P.~S.} \bibnamefont{Julienne}},
  \bibinfo{journal}{Phys. Rev. A} \textbf{\bibinfo{volume}{61}},
  \bibinfo{pages}{022721} (\bibinfo{year}{2000}).

\bibitem[{\citenamefont{Marte et~al.}(2002)\citenamefont{Marte, Volz, Schuster,
  D{\"u}rr, Rempe, van Kempen, and Verhaar}}]{Verhaar02}
\bibinfo{author}{\bibfnamefont{A.}~\bibnamefont{Marte}},
  \bibinfo{author}{\bibfnamefont{T.}~\bibnamefont{Volz}},
  \bibinfo{author}{\bibfnamefont{J.}~\bibnamefont{Schuster}},
  \bibinfo{author}{\bibfnamefont{S.}~\bibnamefont{D{\"u}rr}},
  \bibinfo{author}{\bibfnamefont{G.}~\bibnamefont{Rempe}},
  \bibinfo{author}{\bibfnamefont{E.~G.~M.} \bibnamefont{van Kempen}},
  \bibnamefont{and} \bibinfo{author}{\bibfnamefont{B.~J.}
  \bibnamefont{Verhaar}}, \bibinfo{journal}{Phys. Rev. Lett.}
  \textbf{\bibinfo{volume}{89}}, \bibinfo{pages}{283202}
  (\bibinfo{year}{2002}).

\bibitem[{\citenamefont{Wynar et~al.}(2002)\citenamefont{Wynar, Freeland, Han,
  Ryu, and Heinzen}}]{heinzen02}
\bibinfo{author}{\bibfnamefont{R.}~\bibnamefont{Wynar}},
  \bibinfo{author}{\bibfnamefont{R.~S.} \bibnamefont{Freeland}},
  \bibinfo{author}{\bibfnamefont{D.~J.} \bibnamefont{Han}},
  \bibinfo{author}{\bibfnamefont{C.}~\bibnamefont{Ryu}}, \bibnamefont{and}
  \bibinfo{author}{\bibfnamefont{D.~J.} \bibnamefont{Heinzen}},
  \bibinfo{journal}{Science} \textbf{\bibinfo{volume}{287}},
  \bibinfo{pages}{1016} (\bibinfo{year}{2002}).

\bibitem[{\citenamefont{Xu et~al.}(2003)\citenamefont{Xu, Mukaiyama,
  Abo-Shaeer, Chin, Miller, and Ketterle}}]{ketterle03}
\bibinfo{author}{\bibfnamefont{K.}~\bibnamefont{Xu}},
  \bibinfo{author}{\bibfnamefont{T.}~\bibnamefont{Mukaiyama}},
  \bibinfo{author}{\bibfnamefont{J.~R.} \bibnamefont{Abo-Shaeer}},
  \bibinfo{author}{\bibfnamefont{J.~K.} \bibnamefont{Chin}},
  \bibinfo{author}{\bibfnamefont{D.~E.} \bibnamefont{Miller}},
  \bibnamefont{and} \bibinfo{author}{\bibfnamefont{W.}~\bibnamefont{Ketterle}},
  \bibinfo{journal}{Phys. Rev. Lett.} \textbf{\bibinfo{volume}{91}},
  \bibinfo{pages}{210402} (\bibinfo{year}{2003}).

\bibitem[{\citenamefont{Donley et~al.}(2002)\citenamefont{Donley, Claussen,
  Thompson, and Wieman}}]{wieman02}
\bibinfo{author}{\bibfnamefont{E.}~\bibnamefont{Donley}},
  \bibinfo{author}{\bibfnamefont{N.~R.} \bibnamefont{Claussen}},
  \bibinfo{author}{\bibfnamefont{S.~T.} \bibnamefont{Thompson}},
  \bibnamefont{and} \bibinfo{author}{\bibfnamefont{C.~E.}
  \bibnamefont{Wieman}}, \bibinfo{journal}{Nature}
  \textbf{\bibinfo{volume}{417}}, \bibinfo{pages}{529} (\bibinfo{year}{2002}).

\bibitem[{\citenamefont{Regal et~al.}(2003)\citenamefont{Regal, Ticknor, Bohn,
  and Jin}}]{regal03}
\bibinfo{author}{\bibfnamefont{C.~A.} \bibnamefont{Regal}},
  \bibinfo{author}{\bibfnamefont{C.}~\bibnamefont{Ticknor}},
  \bibinfo{author}{\bibfnamefont{J.~L.} \bibnamefont{Bohn}}, \bibnamefont{and}
  \bibinfo{author}{\bibfnamefont{D.~S.} \bibnamefont{Jin}},
  \textbf{\bibinfo{volume}{424}}, \bibinfo{pages}{47} (\bibinfo{year}{2003}).

\bibitem[{\citenamefont{Herbig et~al.}(2003)\citenamefont{Herbig, Kraemer,
  Mark, Weber, Chin, N{\"a}gerl, and Grimm}}]{grimm03}
\bibinfo{author}{\bibfnamefont{J.}~\bibnamefont{Herbig}},
  \bibinfo{author}{\bibfnamefont{T.}~\bibnamefont{Kraemer}},
  \bibinfo{author}{\bibfnamefont{M.}~\bibnamefont{Mark}},
  \bibinfo{author}{\bibfnamefont{T.}~\bibnamefont{Weber}},
  \bibinfo{author}{\bibfnamefont{C.}~\bibnamefont{Chin}},
  \bibinfo{author}{\bibfnamefont{H.~C.} \bibnamefont{N{\"a}gerl}},
  \bibnamefont{and} \bibinfo{author}{\bibfnamefont{R.}~\bibnamefont{Grimm}},
  \bibinfo{journal}{Science} \textbf{\bibinfo{volume}{301}},
  \bibinfo{pages}{1510} (\bibinfo{year}{2003}).

\bibitem[{\citenamefont{Cubizolles et~al.}(2003)\citenamefont{Cubizolles,
  Bourdel, Kokkelmans, Shlyapnikov, and Salomon}}]{salomon03}
\bibinfo{author}{\bibfnamefont{J.}~\bibnamefont{Cubizolles}},
  \bibinfo{author}{\bibfnamefont{T.}~\bibnamefont{Bourdel}},
  \bibinfo{author}{\bibfnamefont{S.}~\bibnamefont{Kokkelmans}},
  \bibinfo{author}{\bibfnamefont{G.~V.} \bibnamefont{Shlyapnikov}},
  \bibnamefont{and} \bibinfo{author}{\bibfnamefont{C.}~\bibnamefont{Salomon}},
  \bibinfo{journal}{Phys. Rev. Lett.} \textbf{\bibinfo{volume}{91}},
  \bibinfo{pages}{240401} (\bibinfo{year}{2003}).

\bibitem[{\citenamefont{D{\"u}rr et~al.}(2004)\citenamefont{D{\"u}rr, Volz,
  Marte, and Rempe}}]{rempe04}
\bibinfo{author}{\bibfnamefont{S.}~\bibnamefont{D{\"u}rr}},
  \bibinfo{author}{\bibfnamefont{T.}~\bibnamefont{Volz}},
  \bibinfo{author}{\bibfnamefont{A.}~\bibnamefont{Marte}}, \bibnamefont{and}
  \bibinfo{author}{\bibfnamefont{G.}~\bibnamefont{Rempe}},
  \bibinfo{journal}{Phys. Rev. Lett.} \textbf{\bibinfo{volume}{92}},
  \bibinfo{pages}{020406} (\bibinfo{year}{2004}).

\bibitem[{\citenamefont{DiVincenzo}(2000)}]{DiVincenzoFdP}
\bibinfo{author}{\bibfnamefont{D.~P.} \bibnamefont{DiVincenzo}},
  \bibinfo{journal}{Fort. Phys.} \textbf{\bibinfo{volume}{48}},
  \bibinfo{pages}{771} (\bibinfo{year}{2000}).

\end{thebibliography}
\end{document}